\documentclass[prl, pdflatex, twocolumn, superscriptaddress, notitlepage, floatfix]{revtex4-1}
\usepackage{xcolor}
\usepackage{graphicx}
\usepackage{wrapfig}
\usepackage{amsmath}
\usepackage{amssymb}
\usepackage{multirow}
\usepackage{slashed}
\usepackage{physics}
\usepackage[colorlinks=true, linkcolor=blue, citecolor=blue, urlcolor=blue]{hyperref}
\usepackage{booktabs}
\usepackage{bbm}

\newcommand \ua {U(1)_{\mathrm A}}
\newcommand \ml {m_l}
\newcommand \ms {m_s}
\newcommand \tc {T_c}
\newcommand \lda {\lambda}
\newcommand \ro {\rho(\lda, \ml)}
\newcommand \ru[1] {\rho_U(\lda_{#1})}
\newcommand \av[1] {\left\langle{#1}\right\rangle}
\newcommand \U {\mathcal{U}}
\newcommand \nt {N_\tau}
\newcommand \ns {N_\sigma}
\newcommand \cpi {\chi_\pi}
\newcommand \cdl {\chi_\delta}
\newcommand \cdsc {\chi_\mathrm{disc}}

\newcommand \pbp {\langle \bar{\psi}\psi\rangle}
\newcommand{\su}{SU(2)_\mathrm{L}\times SU(2)_\mathrm{R}}

\newcommand{\cf}{\textit{cf.}}

\begin{document}

	

%
\title{Correlated Dirac Eigenvalues and Axial Anomaly in Chiral Symmetric QCD}
\author{H.-T. Ding}
\affiliation{Key Laboratory of Quark and Lepton Physics (MOE) and Institute of
Particle Physics, Central China Normal University, Wuhan 430079, China}
\author{S.-T. Li}
\affiliation{Institute of Modern Physics, Chinese Academy of Sciences, Lanzhou 730000, China}
\affiliation{Key Laboratory of Quark and Lepton Physics (MOE) and Institute of
Particle Physics, Central China Normal University, Wuhan 430079, China}
\author{Swagato Mukherjee}
\affiliation{Physics Department, Brookhaven National Laboratory, Upton, New York 11973, USA}
\author{A. Tomiya}
\affiliation{RIKEN-BNL Research Center, Brookhaven National Laboratory, Upton, New York 11973, USA}
\author{X.-D. Wang}
\affiliation{Key Laboratory of Quark and Lepton Physics (MOE) and Institute of
Particle Physics, Central China Normal University, Wuhan 430079, China}
\author{Y. Zhang}
\email{yuzhang@mails.ccnu.edu.cn}
\affiliation{Key Laboratory of Quark and Lepton Physics (MOE) and Institute of
Particle Physics, Central China Normal University, Wuhan 430079, China}
\begin{abstract}
  We introduce novel relations between the derivatives [$\pdv*[n]{\ro}{\ml}$] of the
  Dirac eigenvalue spectrum [$\ro$] with respect to the light sea quark mass ($\ml$)
  and the $(n+1)$-point correlations among the eigenvalues ($\lda$) of the massless
  Dirac operator. Using these relations we present lattice QCD results for
  $\pdv*[n]{\ro}{\ml}$ ($n=1, 2, 3$) for $\ml$ corresponding to pion masses
  $m_\pi=160-55$~MeV, and at a temperature of about 1.6 times the chiral phase
  transition temperature. Calculations were carried out using (2+1) flavors of highly
  improved staggered quarks with the physical value of strange quark mass, three
  lattice spacings $a=0.12, 0.08, 0.06$~fm, and lattices having aspect ratios $4-9$.
  We find that $\rho(\lda\to0,\ml)$ develops a peaked structure. This peaked
  structure arises due to non-Poisson correlations within the infrared part of the
  Dirac eigenvalue spectrum,  becomes sharper as $a\to0$, and its amplitude is
  proportional to $\ml^2$. We demonstrate that this $\rho(\lda\to0,\ml)$ is
  responsible for the manifestations of axial anomaly in two-point correlation
  functions of light scalar and pseudoscalar mesons. After continuum and chiral
  extrapolations we find that axial anomaly remains manifested in two-point correlation
  functions of scalar and pseudoscalar mesons in the chiral limit.
\end{abstract}
%
%
\maketitle



\emph{Introduction.--}
The Lagrangian of the (2+1)-flavor quantum chromodynamics (QCD) with the physical
value of strange quark mass ($\ms$) and degenerate up and down light quarks possesses
$\su$ chiral symmetry and $\ua$ axial symmetry in the chiral limit of light quark
mass $\ml\to0$.  The chiral symmetry is spontaneously broken in the vacuum and the
$\ua$ symmetry is anomalously broken due to quantum interactions. For the physical
value of $\ml$, the broken chiral symmetry of the QCD vacuum gets approximately
restored through a smooth crossover at a high temperature
$T\simeq156$~MeV~\cite{Aoki:2009sc, Bazavov:2011nk, Bhattacharya:2014ara,
Bonati:2015bha, Bazavov:2018mes, Borsanyi:2020fev},  and for $\ml\to0$ the
restoration takes place via a chiral phase transition at a temperature
$\tc=132^{+3}_{-6}$~MeV~\cite{Ding:2019prx}.

Owing to the asymptotic freedom of QCD, the $\ua$ axial symmetry becomes an exact
symmetry only for $T\to\infty$. However, the nature of the chiral phase transition
crucially depends on how axial anomaly manifests itself in the two-point correlation
functions of light scalar and pseudoscalar mesons for $T\ge\tc$. If the isotriplet
scalar $\delta$ and the isotriplet pseudoscalar $\pi$ remain non-degenerate at
$T\ge\tc$, then the chiral phase transition is expected to be of second order,
belonging to a three-dimensional $O(4)$ universality class~\cite{Pisarski:1983ms}. But if the $\delta$
and $\pi$ become degenerate at $T\ge\tc$, then 
the chiral phase transition can be either first~\cite{Pisarski:1983ms} or
second order~\cite{Butti:2003nu, Pelissetto:2013hqa, Grahl:2014fna}. For the physical
value of $\ml$, the $\delta$ and  $\pi$ remain nondegenerate around the chiral
crossover~\cite{Cheng:2010fe, Bazavov:2012qja, Buchoff:2013nra, Bhattacharya:2014ara,
Bazavov:2019www}. However, what happens for $T\simeq\tc$ as $\ml\to0$ remains an open
question~\cite{Ohno:2012br, Cossu:2013uua, Chiu:2013wwa, Dick:2015twa,
Tomiya:2016jwr, Brandt:2016daq, Suzuki:2020rla, Aoki:2020noz} due to the lack of state-of-the-art
lattice QCD calculations with controlled continuum and chiral extrapolations.

It has been shown that if Dirac eigenvalue spectrum $\ro$ is an analytic function of
$\ml^2$ and $\lda$ then in the chiral limit $\ua$ anomaly will not be manifested in
differences of up to six-point correlation functions of $\pi$ and $\delta$ that can be
connected via a $\ua$ rotation~\cite{Aoki:2012yj}. However, weakly interacting
instanton~\cite{tHooft:1976rip, tHooft:1976snw} gas motivated $\rho \sim \ml^2
\delta(\lda)$ can lead to nondegeneracy of the two-point $\pi$ and $\delta$
correlation functions even as $\ml\to0$~\cite{Bazavov:2012qja}.  While the $\ml^2$
factor naturally arises from the two light fermion determinants,  the
$\delta(\lda)$-like structure is motivated by the limit when the small shift from
zero to the near-zero modes, resulting from the weak interactions among localized
(quasi) instantons and anti-instantons, can be neglected~\cite{Gross:1980br,
Kanazawa:2014cua}.  Lattice QCD studies show that, for the physical values of $\ml$
and for sufficiently high temperatures, the $T$ dependence of a $\ua$-breaking
measure, the topological susceptibility, follows dilute instanton gas approximation
prediction (for a recent review, see~\cite{Lombardo:2020bvn}). However, whether these
findings arise due to an underlying structure of $\rho \sim \ml^2 \delta(\lda)$  and
what happens for $\ml\to0$ have remained unanswered. Some lattice QCD studies have
observed infrared enhancement in $\rho$~\cite{Bazavov:2012qja, Buchoff:2013nra,
Dick:2015twa, Alexandru:2015fxa, Alexandru:2019gdm}, however, whether such enhancements scale as $\ml^2$ as $\ml\to0$ have
not been demonstrated. In other lattice QCD calculations, no infrared enhancement in
$\rho$ was observed~\cite{Cossu:2013uua, Chiu:2013wwa, Tomiya:2016jwr,
Suzuki:2020rla}, showing the importance of controlling lattice artifacts through
continuum extrapolations. On the other hand,  in Ref.~\cite{Kanazawa:2015xna} it was
argued that if $\pi$ and $\delta$ were to remain nondegenerate at $T\ge\tc$,
then chiral symmetry restoration demands non-Poisson correlations among the infrared
eigenvalues.

In this Letter we connect all the above issues: first, by establishing novel relations
between $\pdv*[n]{\rho}{\ml}$ and correlation among the eigenvalues, then by obtaining
$\pdv*[n]{\rho}{\ml}$ from state-of-the-art lattice QCD calculations. Finally, we demonstrate
how the signature of axial anomaly in two-point  $\delta$ and $\pi$ correlation
functions arises as $\ml\to0$.


\emph{$\pdv*[n]{\rho}{\ml}$ and $\ua$ anomaly.--}
For (2+1)-flavor QCD, the Dirac eigenvalue spectrum is given by
\begin{equation}
  \label{eq:ro}
  \begin{split}
    \ro = \frac{T}{V Z[\U]} &
    \int \mathcal{D}[\U] e^{-S_G[\U]} \det\qty[\slashed{D}[\U]+\ms] \\
    & \times \qty(\det\qty[\slashed{D}[\U]+\ml])^2 \ru{} \,.
  \end{split}
\end{equation}
Here, $\ru{} = \sum_j \delta(\lda-\lda_j)$,  $\lda_j$ are the eigenvalues of the
massless Dirac matrix $\slashed{D}[\U]$ for a given background SU(3) gauge field
$\U$,  $V$ is the spatial volume, $S_G[\U]$ is the gauge action, and the partition
function $Z[\U] = \int \mathcal{D}[\U] e^{-S_G[\U]} \det\qty[\slashed{D}[\U]+\ms]
\qty(\det\qty[\slashed{D}[\U]+\ml])^2$. Note that $\ru{}$ does not explicitly depend
on $\ml$, however, $\ml$ dependence enters $\rho$ through the integration over the
gauge fields. Furthermore,
\begin{equation}
  \label{eq:det}
  \begin{split}
    & \det\qty[\slashed{D}[\U]+\ml] = \prod_j \qty(+\mathrm{i}\,\lda_j+\ml) \qty(-\mathrm{i}\,\lda_j+\ml) \\
    & \qquad \qquad \quad
    = \exp \qty( \int_0^\infty \dd{\lda} \ru{} \ln\qty[\lda^2+\ml^2] ) \,.
  \end{split}
\end{equation}
Substituting \autoref{eq:det} in \autoref{eq:ro} and $Z[\U]$ it is straightforward to
obtain $\pdv*[n]{\rho}{\ml}$, \text{e.g.,}
\begin{align}
  \label{eq:dro1}
  & \frac{V}{T} \pdv{\rho}{\ml} = \int_0^\infty \dd{\lda_2}
  \frac { 4 \ml\,  C_2(\lda, \lda_2; \ml)} {  \lda_2^2 + \ml^2 } \,, \\
  \label{eq:dro2}
  \begin{split}
    & \frac{V}{T} \pdv[2]{\rho}{\ml} =
   \int_0^\infty \dd{\lda_2}
    \frac { 4 (\lda_2^2-\ml^2) \,C_2(\lda, \lda_2; \ml) }
    { \qty( \lda_2^2 + \ml^2 )^2}
    \\ & \qquad
    + \int_0^\infty \dd{\lda_2} \dd{\lda_3}
    \frac { (4\ml)^2\, C_3(\lda, \lda_2,\lda_3; \ml) }
    { \qty( \lda_2^2 + \ml^2 ) \qty( \lda_3^2 + \ml^2 ) }
    \,, \qq{with}
  \end{split} \\
  \begin{split}
  \label{eq:Cn}
  &  C_n(\lda_1, \cdots, \lda_n; \ml) =
  \av{\prod_{i=1}^n \qty[ \ru{i} - \av{\ru{i}} ] } .
\end{split}
\end{align}

The difference of the integrated two-point functions, \text{i.e.,} susceptibilities,
of the isotriplet pseudoscalar,
$\pi^i(x)=\mathrm{i}\bar\psi_l(x)\gamma_5\tau^i\psi_l(x)$, and the isotriplet  scalar,
$\delta^i(x)=\bar\psi_l(x)\tau^i\psi_l(x)$, mesons is defined as
\begin{equation}
  \label{eq:cdif}
  \cpi - \cdl = \int \dd[4]{x} \av{ \pi^i(x)\pi^i(0) - \delta^i(x)\delta^i(0) } \,.
\end{equation}
For $T \ge \tc$ owing to the degeneracy of $\pi$ and the isosinglet scalar meson in
the chiral limit~\cite{Bazavov:2012qja}
\begin{equation}
  \label{eq:rel}
  \cpi - \cdl = \cdsc \,,
\end{equation}
where $\cdsc$ is the quark-line disconnected part of the isosinglet scalar meson
susceptibility~\footnote{The term $<\bar\psi\psi>^2$ arises naturally from
$\partial<\bar{\psi}\psi>/\partial m_l$, and trivially vanishes for $T \ge \tc$ as $\ml\to0$.
Following standard convention, this term was subtracted off to cancel the $(\ml/a)^2$
divergence.},
\begin{equation}
  \label{eq:cdsc}
  \cdsc = \frac{T}{V} \int \dd[4]{x} \av{ \qty[ \bar\psi(x)\psi(x)
   - \av{\bar\psi(x)\psi(x)} ]^2 }  \,.
\end{equation}
These $\mathrm{\ua}$ symmetry-breaking measures are related to $\rho$
through~\cite{Bazavov:2012qja, Toublan:2000dn, Kanazawa:2015xna}
\begin{align}
  \label{eq:sus-rho1}
  & \cpi - \cdl = \int_0^\infty \dd{\lda} \frac { 8 \ml^2 \,\rho } { \qty( \lda^2 + \ml^2
  )^2 } \,, \\
  \label{eq:sus-rho2}
  & \cdsc = \int_0^\infty \dd{\lda} \frac { 4 \ml\,\pdv*{\rho}{\ml}} { \lda^2 +\ml^2  } \,.
\end{align}
In the Poisson limit, $C_n$ is given by:  $C_n^{\mathrm{Po}}(\lda_1, \cdots, \lda_n) =
\delta(\lda_1-\lda_2) \cdots \delta(\lda_n-\lda_{n-1})
\av{\qty(\ru{1}-\av{\ru{1}})^n} = \delta(\lda_1-\lda_2) \cdots
\delta(\lda_n-\lda_{n-1}) \av{\ru{1}} + \order{1/N}$, where $2N\propto V/T$ is the total
number of eigenvalues. In this limit,
\begin{align}
  \label{eq:droPo}
  & \qty(\pdv{\rho}{\ml})^{\mathrm{Po}} = \
  \frac{4\ml\rho}{\lda^2+\ml^2} - \frac{V\rho}{TN} \av{\bar\psi\psi} \,,\\
  \label{eq:dro2Po}
  & \qty(\pdv[2]{\rho}{\ml})^{\mathrm{Po}} = \frac{4\rho}{\lda^2+\ml^2} +
  \frac{8\ml^2\rho}{\qty(\lda^2+\ml^2)^2} + \frac{2V^2\rho}{T^2N^2} \av{\bar\psi\psi}^2
  \nonumber \\ & \qquad\qquad
 - \frac{V\rho}{TN} \qty( \frac{8\ml\av{\bar\psi\psi}}{\lda^2+\ml^2} + 2\cpi - \cdl ) \,,
\end{align}
where $\av{\bar\psi\psi}= (T/V)(d\ln Z[\U]/d\ml)$. In the chiral limit, this
leads to $\cdsc^\mathrm{Po} = 2(\cpi-\cdl)$, in clear violation of the chiral
symmetry restoration condition in \autoref{eq:rel}, unless both sides of the equation
trivially vanish.


\emph{Lattice QCD calculations.--}
Lattice QCD calculations were carried out at $T \approx   205$~MeV $\approx1.6\tc$ for
$(2+1)$-flavor QCD using the highly improved staggered quarks and the tree-level
Symanzik gauge action, a setup extensively used by the HotQCD
Collaboration~\cite{Bazavov:2011nk, Bazavov:2014pvz, Bazavov:2012jq, Bazavov:2017dus,
Bazavov:2018mes}. The $\ms$ was tuned to its physical value and three lattice
spacings $a = (T\nt)^{-1} = 0.12, 0.08, 0.06$~fm, corresponding to lattice temporal
extents $\nt=8, 12, 16$, were used~\cite{Bazavov:2019www}. Calculations were done
with $\ml = \ms/20, \ms/27, \ms/40, \ms/80, \ms/160$ that correspond to $m_\pi \simeq
160, 140, 110, 80, 55$~MeV, respectively. The spatial extents ($\ns$) of the lattices
were chosen to have aspect ratios in the range of $\ns/\nt=4-9$.  The gauge field
configurations were generated using the rational hybrid Monte Carlo
algorithm~\cite{Clark:2004cp,Bazavov:2010ru}. Gauge configurations from every
10$\mathrm{th}$ molecular dynamics trajectory of unit length were saved to carry out
various measurements. $\rho$ and $C_n$ were computed by measuring  $\ru{}$ over the
entire range of $\lda$ using the Chebyshev filtering technique combined with the
stochastic estimate method~\cite{Ding:2020eql, Giusti:2008vb, Cossu:2016eqs,
Fodor:2016hke, Ding:2020hxw} on $\sim2000$ configurations. Orders of the Chebyshev
polynomials were chosen to be $(1-5)\times10^5$ and 24 Gaussian stochastic sources
were used. Measurements of $\cdsc$ and $\cpi-\cdl$ were done by inverting the light
fermion matrix using $50$ Gaussian random sources on $2000-10000$
configurations~\footnote{$\cpi$ was obtained through the Ward identity
$\ml\cpi=<\bar\psi\psi>$~\cite{Ding:2020hxw,Kilcup:1986dg} and $\cdl$ is the
connected chiral susceptibility~\cite{Buchoff:2013nra}. We checked that these results
were reproduced, within errors, by the susceptibilities of the independently computed
$\pi$ and $\delta$ 2-point correlation functions.}.


%
\begin{figure*}[!htp]
  \centering
	\includegraphics[width=0.32\textwidth, height=0.17\textheight]{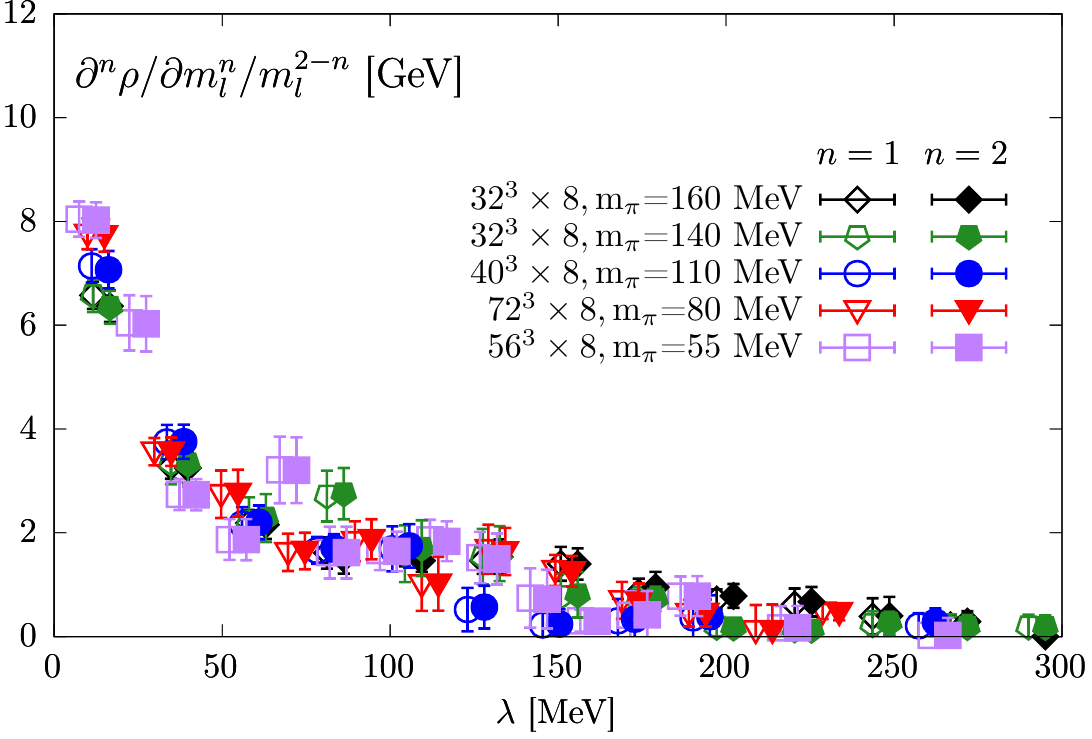}
	\includegraphics[width=0.32\textwidth, height=0.17\textheight]{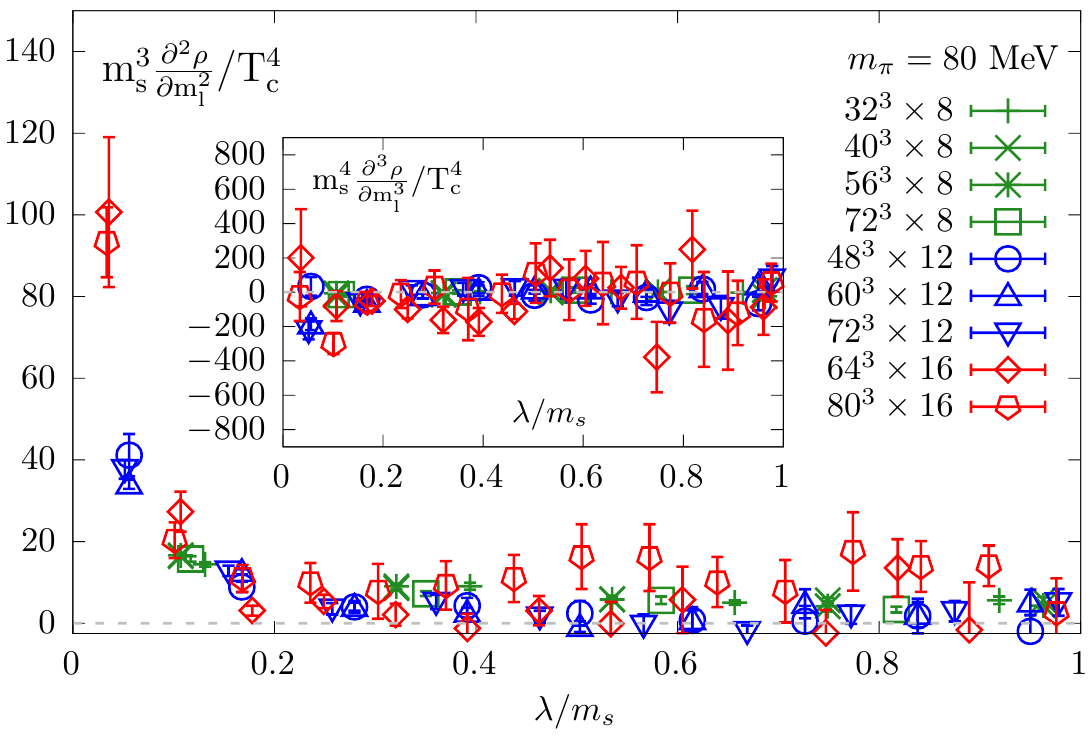}
         \includegraphics[width=0.32\textwidth,height=0.17\textheight]{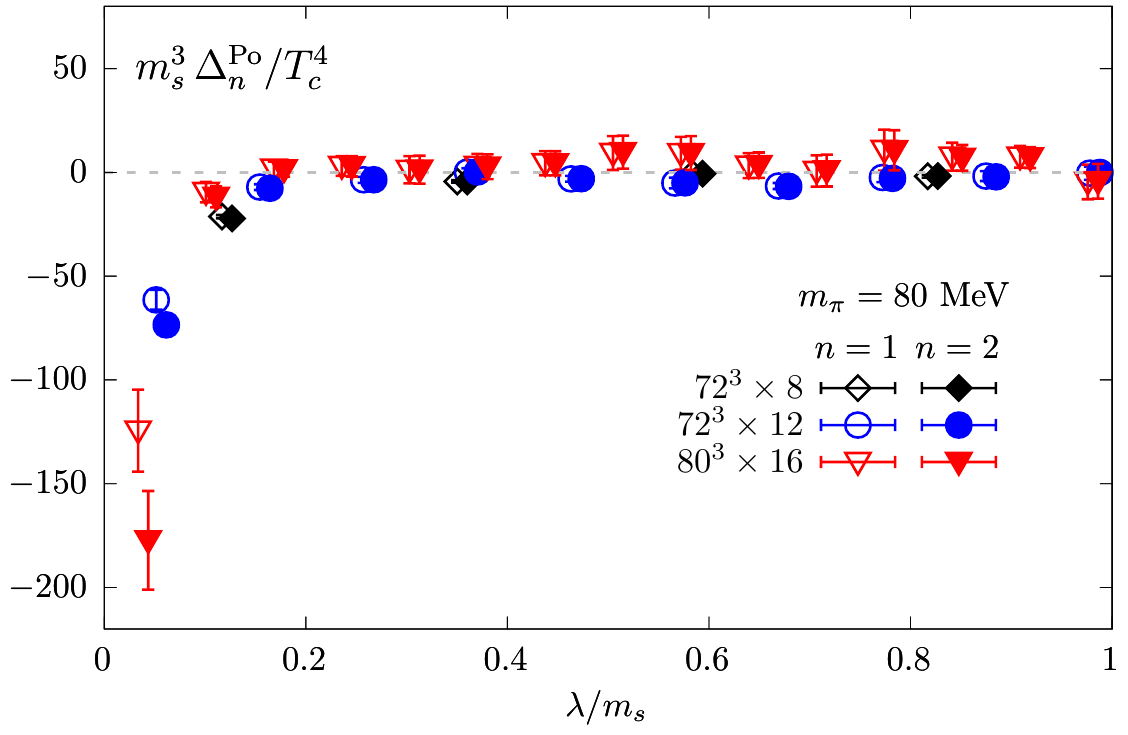}
	\caption{Left: Light sea quark mass dependence of $m_l^{-1}\pdv*{\ro}{\ml}$ (open
	symbols) and $\pdv*[2]{\ro}{\ml}$ (filled symbols) using $\nt=8$ lattices. Middle:
	Lattice spacing and volume dependence of $\pdv*[2]{\ro}{\ml}$ and
	$\pdv*[3]{\ro}{\ml}$ (inset) for $m_\pi=80$~MeV. Right: The differences,
	$\Delta_n^\mathrm{Po} = \ml^{n-2} \qty[ \pdv*[n]{\rho}{\ml} -
	(\pdv*[n]{\rho}{\ml})^\mathrm{Po} ]$ [\text{cf.} \autoref{eq:droPo} and
	\autoref{eq:dro2Po}], for $m_\pi=80$~MeV and three lattice spacings. In all cases,
	results are obtained at $T\approx205$~MeV and the filled symbols have been slightly
	shifted horizontally for visibility. 
  }
	\label{fig:rho}
\end{figure*}
\emph{Results.--}
\autoref{fig:rho} (left) shows the $\ml$ dependence of $m_l^{-1}\pdv*{\rho}{\ml}$ and
$\pdv*[2]{\rho}{\ml}$ at $T\approx1.6\tc$, obtained for lattices with $\nt=8$ and the
largest available $\ns$ for that $\ml$. We observe that $\ml^{-1}(\pdv*{\rho}{\ml})$
and $\pdv*[2]{\rho}{\ml}$ are almost identical and independent of $\ml$. Also,
$\ml^{-1}\pdv*{\rho}{\ml}$ and $\pdv*[2]{\rho}{\ml}$ are peaked at $\lda\to0$ and drop
rapidly toward zero for $\lda/T\gtrsim1$. \autoref{fig:rho} (middle) depicts the
lattice spacing and volume dependence of $\pdv*[2]{\rho}{\ml}$ and
$\pdv*[3]{\rho}{\ml}$ for $m_\pi=80$~MeV. To compare these quantities across
different lattice spacings we multiply with the appropriate powers of $\ms$ to make
them renormalization group invariant and make them dimensionless by rescaling
with appropriate powers of $\tc=132$~MeV. We see that the  peaked structure in
$\pdv*[2]{\rho}{\ml}$ at $\lda\to0$ becomes sharper as $a\to0$, and shows little
volume dependence(see Supplemental Material~\footnote{ See Supplemental Material  for the technical details of this study, which includes Refs.~\cite{Bazavov:2011nk, Bazavov:2014pvz, Bazavov:2012jq, Bazavov:2017dus,
		Bazavov:2018mes,Kanazawa:2015xna,Ramos:2015baa,Giusti:2008vb,Patella:2012da,Cossu:2016eqs,Fodor:2016hke,Ding:2020hxw,Ding:2020eql,Itou:2014ota,deForcrand:2017cja,NapoliPS13,PowerMethod:Saad92}.  We also found that even when the same bin size in $\lambda/m_s$ is used for all three lattice spacings the infrared peak structure persists as continuum limit is approached, see~\autoref{fig:sup_fig1middle2} of the Supplemental Material. }). Moreover, within errors, $\pdv*[3]{\rho}{\ml}$ are found to be
consistent with zero in all the cases. 
The findings
$\ml^{-1}\pdv*{\rho}{\ml}\approx\pdv*[2]{\rho}{\ml}$ and
$\pdv*[3]{\rho}{\ml}\approx0$ show that the peaked structure
$\rho(\lda\to0,\ml\to0)\propto\ml^2$. In \autoref{fig:rho} (right) we show the difference
$\Delta_n^\mathrm{Po} = \ml^{n-2} \qty[ \pdv*[n]{\rho}{\ml} -
(\pdv*[n]{\rho}{\ml})^\mathrm{Po} ]$ ($n=1,2$), with the Poisson approximations for
$\pdv*[n]{\rho}{\ml}$ as defined in \autoref{eq:droPo} and \autoref{eq:dro2Po}. The
fact $\Delta_n^\mathrm{Po}<0$ shows that the repulsive non-Poisson correlation
within the small $\lda$ gives rise to the $\rho(\lda\to0)$ peak.

\begin{figure}[!htp]
  \centering
	\includegraphics[width=0.32\textwidth, height=0.17\textheight]{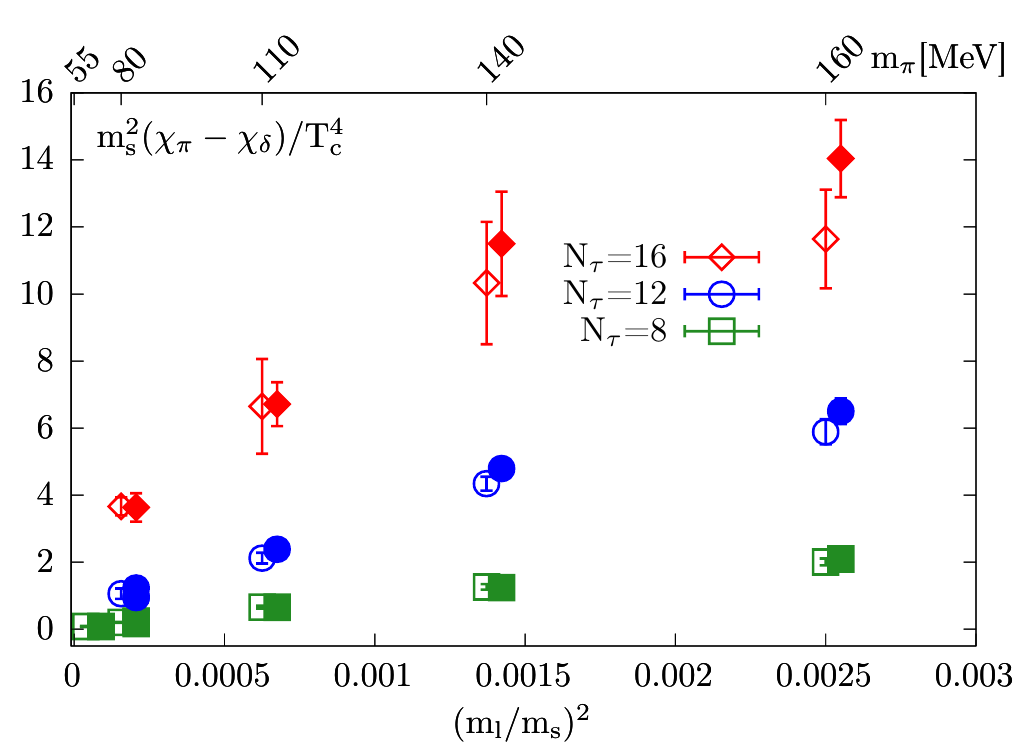}
	\includegraphics[width=0.32\textwidth, height=0.17\textheight]{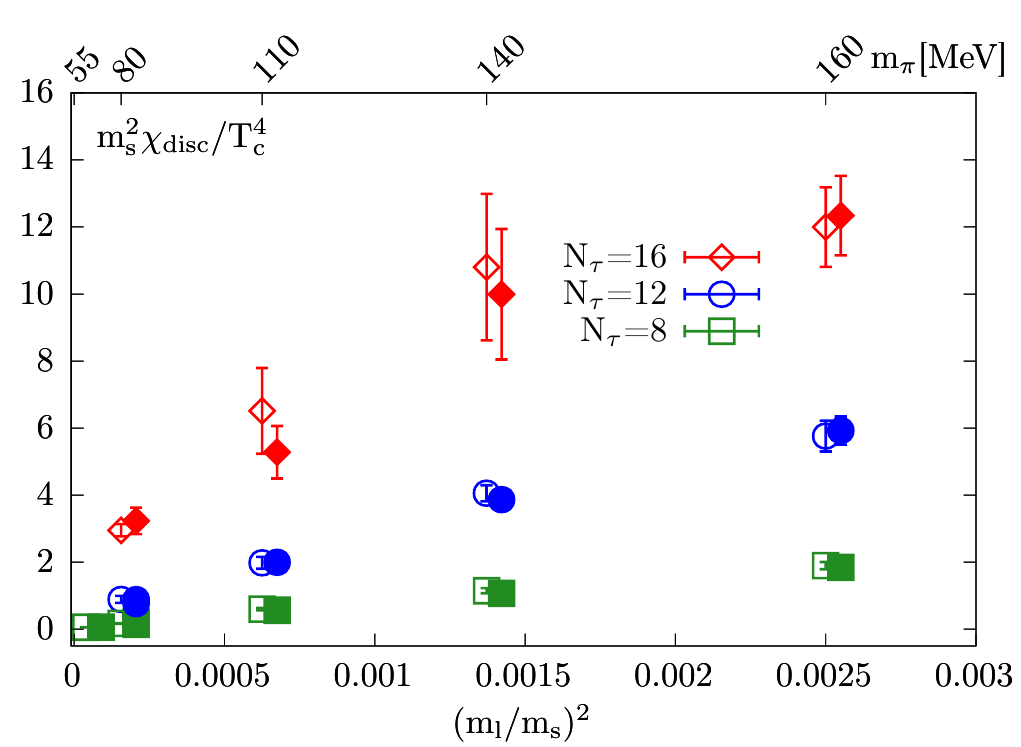}
	\caption{Comparisons of direct measurements (open symbols) of $\cpi-\cdl$ (top) and
	$\cdsc$ (bottom) with those reconstructed (filled symbols, slightly shifted
	horizontally for visibility) from $\rho$ [\text{cf.} \autoref{eq:sus-rho1}] and
	$\pdv*{\rho}{\ml}$ [\text{cf.} \autoref{eq:sus-rho2}], respectively. The results
	are shown for all values of light quark masses and lattice spacings at
	$T\approx205$~MeV.}
  \label{fig:Comparison}
\end{figure}

In \autoref{fig:Comparison} we show that $\rho$ and $\pdv*{\rho}{\ml}$ reproduce
directly measured $\cpi-\cdl$ and $\cdsc$ using \autoref{eq:sus-rho1} and
\autoref{eq:sus-rho2}, respectively. The numerical integrations in $\lambda$ were performed using the rectangle 
method, where the largest value of $\lambda$ 
was estimated using the power method and the statistical error of integration was obtained using the jackknife method.
Since we saw very mild volume dependence in all
the  quantities, we only present results from the largest available volume for each
$\nt$ and $\ml$. We checked that only the infrared $\lda/T\lesssim1$ parts of $\rho$
and $\pdv*{\rho}{\ml}$ are needed for the reproductions of $\cpi-\cdl$ and $\cdsc$,
within errors, for all $\nt$ and $\ml$. Additionally, we checked that once the
bin-size of $\lda$ in the numerical integration of \autoref{eq:sus-rho1} is chosen to
reproduce directly measured $\cpi-\cdl$, the same bin size automatically reproduces
$\cdsc$ and $\av{\bar\psi\psi}$ without any further tuning. We observe that both
$\cdsc$ and $\cpi-\cdl$ are linear in $\ml^2$ for all lattice spacings and especially
for $m_\pi\lesssim 140$~MeV; this is in accord with the expectation $Z[\U]$ is an
even function of $\ml$ for $T\ge\tc$ due to the restoration of the $Z(2)$ subgroup of
$\su$.
\begin{figure}[!htp]
  \centering
	\includegraphics[width=0.32\textwidth, height=0.17\textheight]{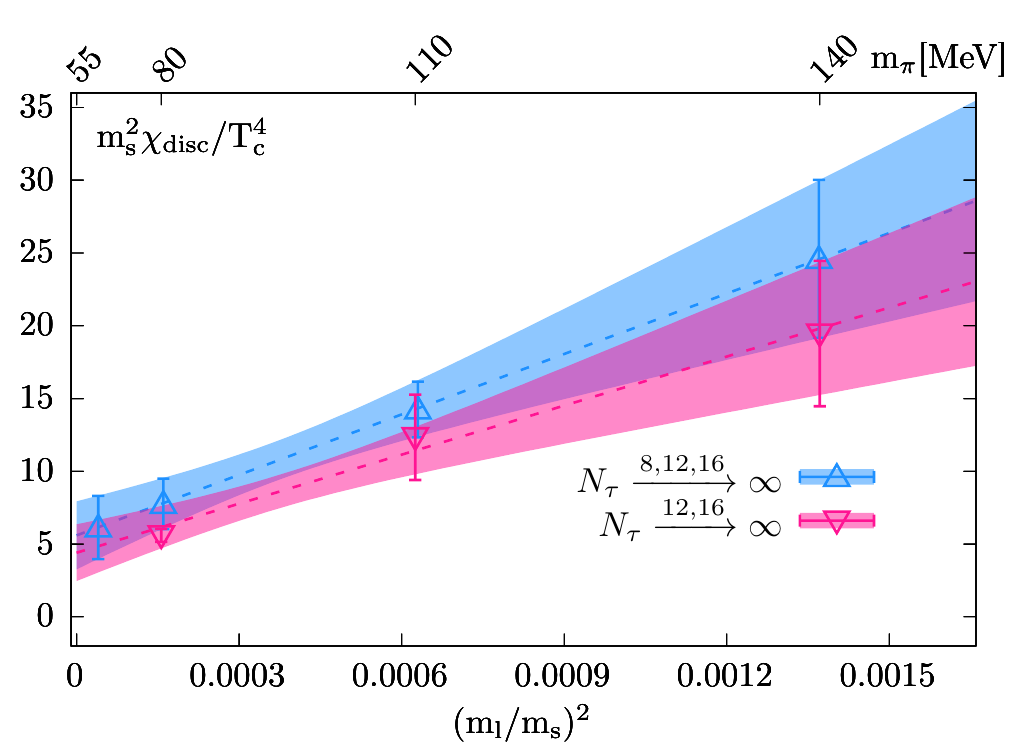}
	\includegraphics[width=0.32\textwidth, height=0.17\textheight]{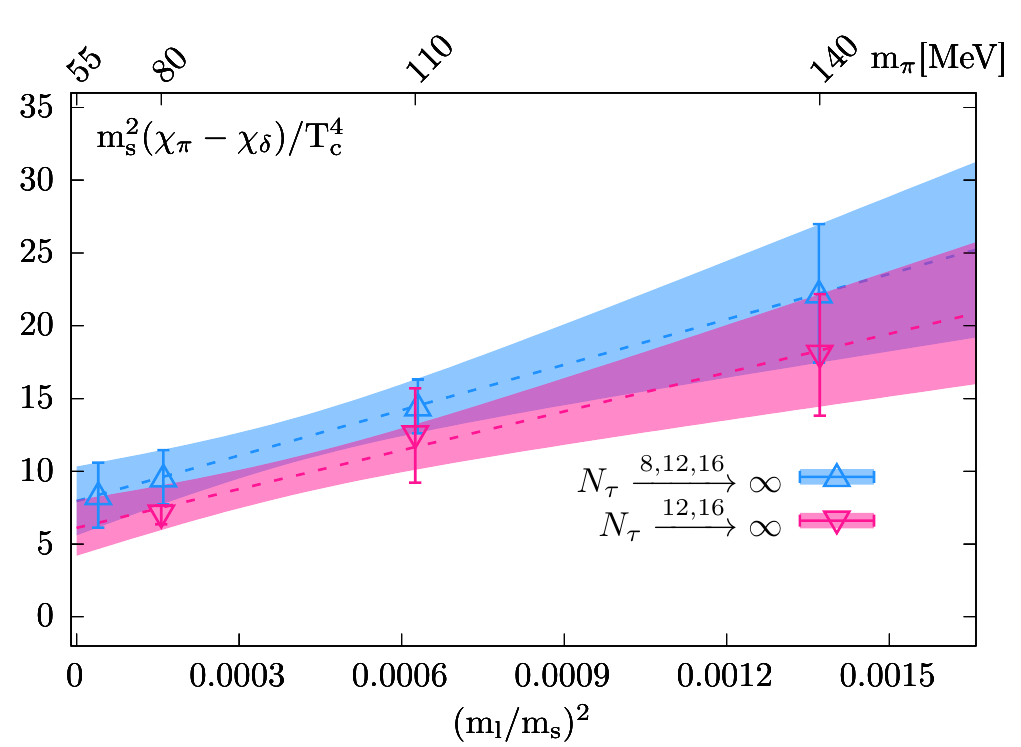}
	\caption{Continuum and chiral extrapolated results for $\cdsc$ (top) and
	$\cpi-\cdl$ (bottom) at $T\approx205$~MeV. See text for details.}
  \label{fig:cont_chiral_chi}
\end{figure}

In \autoref{fig:cont_chiral_chi} we show the continuum and chiral extrapolated
results for $\cdsc$ and $\cpi-\cdl$. Using all the data for $\nt=8, 12, 16$ and
$m_\pi\le140$~MeV, we performed a joint $a, \ml \to 0$ extrapolation of the form
$\cdsc(a,\ml) = \cdsc(0,0) +a_1/\nt^2 + a_2/\nt^4 + \qty(\ml/\ms)^2\qty[ b_0+
b_1/\nt^2 + b_2/\nt^4 ]$. Fits were performed on each bootstrap sample of the data
set. The bootstrap samples were created by randomly choosing data from Gaussian
distributions with means equal to the average values and variances equal to the
1-$\sigma$ errors  of the directly measured $\cdsc$. We chose the median value of the
resulting bootstrap distribution as the final result (depicted by the upward
triangles) and the 68\% percentiles confidence interval of the resulting distribution
as the errors on the final results (the band labeled by
$N_\tau\xrightarrow{8,12,16}\infty$). Since we used the so-called rooted-staggered
formulation~\cite{Bernard:2004ab, Bernard:2006vv, Sharpe:2006re,Kilcup:1986dg} for
our (2+1)-flavor lattice QCD, we also checked that the same $\cdsc(0,0)$ is obtained
within errors by first carrying out the $a\to0$ extrapolations for each $\ml$ and
then performing the $\ml\to0$ extrapolation using the $a\to0$ extrapolated results.
For this purpose, we used the $\nt=12, 16$ data for each of $\ml = \ms/27, \ms/40,
\ms/80$ to obtain $\cdsc(0,\ml)$ by fitting to the ansatz $\cdsc (a,m_{l})=\cdsc
(0,m_{l})+d_{1}/N_{\tau}^{2}$. Then the chiral extrapolation was carried out using
$\cdsc(0,\ml) = \cdsc(0,0) + d_2(\ml/\ms)^2$ based on the continuum estimates of
$\cdsc(0,\ml)$. These extrapolations were done by using the same bootstrap procedure
described before and the final results are indicated with the label
$N_\tau\xrightarrow{12,16}\infty$. Exactly the same procedures were followed also
for $\cpi-\cdl$ to obtain its continuum and chiral extrapolated values. After
carrying out continuum and chiral extrapolations we find that \autoref{eq:rel} is
satisfied within errors, and $\cdsc$ and $\cpi-\cdl$ are nonvanishing at a confidence level above 95\%.

\emph{Conclusions.---}
In this Letter we establish relations between $\pdv*[n]{\rho}{\ml}$ and $C_{n+1}$. To
the best of our knowledge these relations are new in the literature. Based on these
relations, for the first time, we present direct computations of
$\pdv*[n]{\rho}{\ml}$ employing state-of-the-art lattice QCD techniques. The results
presented in this Letter led us to conclude that, in chiral symmetric (2+1)-flavor QCD at
$T\approx1.6\tc$, (i) $\rho(\lda\to0,\ml)$ develops a peaked structure due to
repulsive non-Poisson correlations within small $\lda$; the peak becomes sharper as
$a\to0$, and its amplitude is $\propto\ml^2$. (ii) The underlying presence of this
$\rho(\lda\to0,\ml)$ leads to manifestations of $\ua$ anomaly in $\cpi-\cdl$ and
$\cdsc$. (iii) Axial anomaly remains manifested in $\cpi-\cdl$ and $\cdsc$ even in
the chiral limit. These suggest that for $T\gtrsim1.6\tc$ the microscopic origin of
axial anomaly is driven by the weakly interacting (quasi)instanton gas motivated
$\rho(\lda\to0,\ml\to0)\sim\ml^2\delta(\lda)$, and the chiral phase transition in
(2+1)-flavor QCD is of the three-dimensional $O(4)$ universality class.

The above conclusions are based on the continuum extrapolated lattice QCD
calculations using the (2+1) flavors of staggered fermions. Confirmations of these
continuum extrapolated results using other fermion actions, especially using chiral
fermions, are needed in future. Even in those future calculations it will be very
difficult to directly identify a structure like $\ml^2\delta(\lda)$ in $\rho$ itself
as $\ml\to0$. The formalism developed and techniques presented in this Letter for
directly accessing $\pdv*[n]{\rho}{\ml}$ will be essential for those future studies
too. The same or similar formalism also may have many potential applications beyond
the present physics problem: few plausible examples testing the predictions of random
matrix theory~\cite{Osborn:1998nf, Toublan:2000dn, Christensen:2014raa},
determination of strong coupling constant using Dirac eigenvalue
spectrum~\cite{Nakayama:2018ubk},  determinations of mass anomalous dimensions in
different theories~\cite{Patella:2011jr, Patella:2012da, Cichy:2013eoa, Cheng:2013eu,
Karthik:2020shl}, \text{etc}.


%
This material is based upon work supported by the National Natural Science
Foundation of China under Grants No. 11775096, No. 11535012, and No. 11947237; the U.S.
Department of Energy, Office of Science, Office of Nuclear Physics through the Award No. DE-SC0012704; the U.S. Department of Energy,
Office of Science, Office of Nuclear Physics and Office of Advanced Scientific
Computing Research within the framework of Scientific Discovery through Advance
Computing (SciDAC) award Computing the Properties of Matter with Leadership Computing
Resources; and RIKEN Special Postdoctoral Researcher program and JSPS KAKENHI Grant No. JP20K14479.
Computations for this work were carried out on the GPU clusters of the Nuclear
Science Computing Center at Central China Normal University (NSC$^3$), Wuhan, China,
and facilities of the USQCD Collaboration, which are funded by the Office of Science
of the U.S. Department of Energy.
For generating the gauge configurations, the HotQCD software suite was used, and the
eigenvalue measurement code was developed also based on the same software suite. We
are indebted to the HotQCD Collaboration for sharing their software suite with us.

\bibliographystyle{apsrev4-1.bst}
\bibliography{ref.bib}
%



\begin{widetext}
	\section{\large Supplemental Materials}
	We provide supplemental materials in the sequence according to the contents in the main material.

\section{I. $\partial^{n}\rho/m_{l}^n$ and $U(1)_A$ anomaly}
\subsection{IA. Quantities related to $\rho$ and $\pdv*[2]{\rho}{m_l}$}
The two-flavor light quark chiral condensate $\pbp$ is related to the Dirac eigenvalue spectrum $\rho$ as follows
\begin{equation}
\pbp=\int_0^{\infty}\frac{4m_{l}\,\rho}{\lambda^2 + m_{l}^2}\,\mathrm{d}\lambda\,, \label{eq:sup_pbp_rho}
\end{equation}
and it can also be expressed in terms of fermion matrix inverse $M^{-1}$
\begin{equation}
\pbp =\frac{T}{V} \,\frac{\mathrm{d} \,\mathrm{ln}Z}{\mathrm{d}\, m_l} = \frac{2T}{V} \,\mathrm{Tr} (\slashed{D}+m_l)^{-1} \equiv \frac{2T}{V}\,\mathrm{Tr} M^{-1},
\label{eq:sup_pbp_M}
\end{equation}
where $M$ is the single-flavor fermion matrix. In our lattice QCD simulations staggered fermions are adopted and we thus deal with a 4-flavor fermion matrix ${M_{stag}}$. By utilizing the commonly used fourth-root technique~\cite{Bazavov:2011nk, Bazavov:2014pvz, Bazavov:2012jq, Bazavov:2017dus,
	Bazavov:2018mes} we have $\mathrm{Tr}M^{-1}\equiv\frac{1}{4}\mathrm{Tr}M_{stag}^{-1}$. The same applies to $\rho$, i.e. $\rho\equiv \rho_{stag}/4$ through our paper.

	We define a quantity $\chi_2$ which can be related to the second order derivative of $\rho$ with respect to $m_l$ as follows
\begin{equation}
\chi_2 = \int_{0}^{\infty}{\rm d}\lambda \,\frac{4m_{l}\,\partial^2\rho/\partial m_l^2}{\lambda^2 + m_{l}^2}\,.
\label{eq:sup_chi2_rho}
\end{equation}
The above quantity can also be evaluated in terms of $M^{-1}$,
\begin{equation}
\begin{aligned}
\chi_2 & = \frac{4T}{V} \,\bigg( 2\,\qty(\left\langle\left({\rm Tr} M^{-1}\right)^{3}\right\rangle 
+ 2\left\langle {\rm Tr} M^{-1}\right\rangle^3 -3\left\langle{\rm Tr} M^{-1} \right\rangle \left\langle\left({\rm Tr} M^{-1}\right)^{2}\right\rangle  ) \\
& \quad\quad\quad\quad+  \left\langle{\rm Tr} M^{-1}\right\rangle \left\langle{\rm Tr} M^{-2}\right\rangle - \left\langle {\rm Tr} M^{-2} {\rm Tr} M^{-1} \right\rangle\bigg)\,.
\end{aligned}
\label{eq:sup_chi2_M}
\end{equation}

\subsection{IB. Expressions for $\pdv*[n]{\rho}{\ml}$ and $C_{n+1}$ with n=1,2,3 }
One can work out the expressions for higher order correlation functions and $m_l$ derivatives of $\rho$ according to procedures described in the main material
(\cf~\autoref{eq:ro},~\autoref{eq:det},~\autoref{eq:dro1},~\autoref{eq:dro2} and~\autoref{eq:Cn})
 Here for demonstration we show expressions for 
up to third derivatives of $\rho$ with respect to $m_l$, and up to four point correlation functions $C_4$ as well as $\cdsc$ and $\chi_2$ in the Poisson limits.

The third derivative of $\rho(\lambda,m_{l})$ with respect to the light quark mass can be expressed as follows
\begin{equation}
\begin{aligned}
\frac{V}{T}\,\frac{\partial^3{\rho(\lambda,m_l)}}{\partial m_{l}^3} 
& = \int_{0}^{\infty}  d\lambda_3   \int_{0}^{\infty}  d\lambda_2
\int_{0}^{\infty}  d\lambda_1   \frac{(4m_{l})^3C_4(\lambda,\lambda_1,\lambda_2,\lambda_3; m_{l})}{(\lambda_3^2 + m_{l}^2) (\lambda_2^2 + m_{l}^2) (\lambda_1^2 + m_{l}^2) } \\  
& +  \int_{0}^{\infty}  d\lambda_2 
\int_{0}^{\infty}  d\lambda_1 \frac{48m_{l}(\lambda_2^2 - m_{l}^2) C_3(\lambda,\lambda_1,\lambda_2;m_{l}) }{(\lambda_2^2 + m_{l}^2)^2(\lambda_1^2 + m_{l}^2)}\\
& + \int_{0}^{\infty}d\lambda_1  \frac{8m_{l} (m_{l}^2 - 3\lambda_1^2) C_2(\lambda,\lambda_1;m_l)}{(\lambda_1^2 +m_{l}^2 )^3}.   
\end{aligned}
\end{equation}
where $C_2$ and $C_3$ are two-point and three-point correlation functions, respectively, as mentioned in the main text, and $C_4$ is the four point correlation function.

In the case that number of eigenvalues among gauge ensembles is Poisson distributed, the correlation functions are reduced to
\begin{align}
C_2^{\mathrm Po}(\lambda_1, \lambda_2;m_l) & =\frac{V}{T}\qty( \delta(\lambda_1 - \lambda_2)\rho(\lambda_1,m_l) - \frac{V}{TN} \rho(\lambda_1,m_l) \rho(\lambda_2,m_l)),\\ 
\begin{split}
C_3^{\mathrm{Po}}(\lambda_1, \lambda_2, \lambda_3; m_l) & =\frac{V}{T}\bigg(  \delta(\lambda_2 - \lambda_1)\delta(\lambda_3-\lambda_1)\rho(\lambda_1,m_l)\\
&	- \frac{V}{TN}\Big(\delta(\lambda_2 - \lambda_1) \rho(\lambda_1,m_l) \rho(\lambda_3,m_l) + \delta(\lambda_3 - \lambda_2)\rho(\lambda_1,m_l) \rho(\lambda_2,m_l) \\
&\qquad \,\,~~+\delta(\lambda_3 - \lambda_1)\rho(\lambda_1,m_l) \rho(\lambda_2,m_l) \Big)\\
& + \Big(\frac{V}{TN}\Big)^2 \,2 \rho(\lambda_1,m_l)\rho(\lambda_2,m_l) \rho(\lambda_3,m_l)\bigg)\,,
\end{split}\\
\begin{split}
C_4^{\mathrm{Po}}(\lambda_1, \lambda_2, \lambda_3, \lambda_4;m_l) & =\frac{V}{T}\bigg( \delta(\lambda_2 - \lambda_1)\delta(\lambda_3-\lambda_1)\delta(\lambda_4-\lambda_1)\rho(\lambda_1,m_l)\\
&-\frac{V}{TN}\Big( \delta(\lambda_2 - \lambda_1)\delta(\lambda_3 - \lambda_1)\rho(\lambda_1,m_l)\rho(\lambda_4,m_l)+\delta(\lambda_1 - \lambda_4) \delta(\lambda_2 - \lambda_1) \rho(\lambda_1,m_l)\rho(\lambda_3,m_l)  \\
& \qquad \,\,~~ + \delta(\lambda_3-\lambda_4)\delta(\lambda_2-\lambda_1) \rho(\lambda_1,m_l)\rho(\lambda_3,m_l) + \delta(\lambda_1 - \lambda_4) \delta(\lambda_3-\lambda_2) \rho(\lambda_1,m_l)\rho(\lambda_2,m_l)\\
& \qquad \,\,~~  +\delta(\lambda_2- \lambda_4)\delta(\lambda_3-\lambda_2) \rho(\lambda_1,m_l)\rho(\lambda_2,m_l) + \delta(\lambda_3 - \lambda_1)\delta(\lambda_2 - \lambda_4) \rho(\lambda_1,m_l) \rho(\lambda_2,m_l) \\
& \qquad \,\,~ + \delta(\lambda_1 - \lambda_4)\delta(\lambda_3-\lambda_1)\rho(\lambda_1,m_l)\rho(\lambda_2,m_l) \Big)\\
& + 2\left(\frac{V}{TN}\right)^2 \Big(\delta(\lambda_2-\lambda_1) \rho(\lambda_1,m_l) \rho(\lambda_3,m_l) \rho(\lambda_4,m_l)+ \delta(\lambda_3-\lambda_2) \rho(\lambda_1,m_l)\rho(\lambda_2,m_l)\rho(\lambda_4,m_l)\\
& \qquad \qquad \,\,\,~~~~~+  \delta(\lambda_3-\lambda_1)\rho(\lambda_1,m_l) \rho(\lambda_2,m_l) \rangle\rho(\lambda_4,m_l) +   \delta(\lambda_1- \lambda_4)\rho(\lambda_1,m_l)\rho(\lambda_2,m_l)\rho(\lambda_3,m_l)\\
& \qquad \qquad \,\,\,~~~~~+   \delta(\lambda_2 - \lambda_4) \rho(\lambda_1,m_l)\rho(\lambda_2,m_l)\rho(\lambda_3,m_l) +\delta(\lambda_3 - \lambda_4) \rho(\lambda_1,m_l)\rho(\lambda_2,m_l) \rho(\lambda_3,m_l) \Big)\\
& - 6\left(\frac{V}{TN}\right)^3 \rho(\lambda_1,m_l) \rho(\lambda_2,m_l)\rho(\lambda_3,m_l)\rho(\lambda_4,m_l)\bigg)\,.
\end{split}
\end{align}
As briefly mentioned in the main material $N$ is the number of chiral pairs of Dirac eigenvalues $\{\pm i\lambda_n\}_{n=1}^{N}$ and it equals to the half number of the lattice sites, $N=N_\sigma^3N_\tau/2$. The detailed derivation of $C_2^{\mathrm{Po}}$ can be found, e.g. in the appendix of Ref.~\cite{Kanazawa:2015xna} and $C_n^{\mathrm{Po}}$ with $n\geq3$ can be obtained following the same procedures.

The resulting expressions of $\chi_{disc}$, $\chi_2$ and ${\partial^3{\rho(\lambda,m_l)}}/{\partial m_{l}^3}$ in the Poisson limit are listed as follows
\begin{align}
\begin{split}
\qty(\chi_{\rm {disc}})^{\mathrm{Po}} & =  2(\chi_{\pi} - \chi_{\delta}) - \frac{V}{TN}\langle \bar{\psi}\psi \rangle^2\,, 
\end{split}\\
\begin{split}
\qty(\chi_2)^{\mathrm{Po}}		& = \int_{0}^{\infty}{\rm d}\lambda\frac{16m_l\,\rho(\lambda,m_{l})(\lambda^2 +3m_{l}^2)}{(\lambda^2 + m_{l}^2)^3} - \frac{V}{TN}  \langle \bar{\psi}\psi \rangle (6\chi_\pi -5 \chi_\delta)  +  2\qty(\frac{V}{TN})^2 \langle \bar{\psi}\psi \rangle^3, 
\end{split}\\
\begin{split}
\qty(\frac{\partial^3{\rho(\lambda,m_l)}}{\partial m_l^3})^{\mathrm{Po}}  &= \frac{24m_l\,\rho(\lambda,m_l)}{(\lambda^2 + m_l^2)^2 } \\&- 12\,\frac{V}{TN}\,\rho(\lambda,m_l)\Bigg(\int_{0}^{\infty}{\rm d}\lambda_1 \frac{2m_l\,\rho(\lambda_1,m_l)}{(\lambda_1^2+m_l^2)^2} -  \langle\bar{\psi}\psi\rangle \frac{3m_{l}^2 + \lambda^2}{(\lambda^2 + m_l^2)^2} 
 -\frac{m_l(2\chi_{\pi} - \chi_{\delta})}{\lambda^2 + m_l^2}\Bigg)\\ 
 &+ 6\,\left(\frac{V}{TN}\right)^2
\rho(\lambda,m_l)\langle\bar{\psi}\psi\rangle  \left(\frac{4m_l}{\lambda^2 + m^2}\langle\bar{\psi}\psi\rangle +  (2\chi_\pi - \chi_\delta) \right) \\
&  - 6\,\left(\frac{V}{TN}\right)^3 \rho(\lambda,m_l)\langle\bar{\psi}\psi \rangle^3 \,.
\end{split}
\label{eq:sup_obs_Po}
\end{align}
 $\pbp$ vanishes in the chiral limit at $T>T_c$, and this leads to $\qty(\chi_{\rm {disc}})^{\mathrm{Po}}  =  2(\chi_{\pi} - \chi_{\delta}) $ as mentioned in the main material.

\section{II. Lattice QCD calculations}
	\subsection{II A. Data, Statistics and parameters used in the lattice setup}
	\label{sec:sup_setup}
		In this subsection we show the simulation parameters as well as statistics in~\autoref{sup:table_setup}. We also list the direct measurements of $\mathrm{m_{s}^2\cdsc/T_{c}^{4}}$ and $\mathrm{m_{s}^2(\chi_{\pi} - \chi_{\delta})/T_{c}^{4}}$. 	In~\autoref{fig:sup_top_history} we show the time history of topological charge obtained from the finest lattices we have, $80^3\times16$ lattices with $m_\pi=80$ MeV and datasets with the smallest quark mass, $56^3\times8$ with $m_\pi=55$ MeV.  

	\begin{table}[htb]
	\centering
	\begin{tabular}{*{10}{c}}
		\toprule \hline
		\toprule  
		\multirow{2}*{$\beta$}  & \multirow{2}*{$am_s$} & \multirow{2}*{$am_l$} & \multirow{2}*{$m_{\pi}$[MeV]} & \multirow{2}*{$N_{\sigma}^3\times N_{\tau}$} & \multicolumn{2}{c}{Chebyshev} & \multicolumn {3}{c}{Direct}  \\  
		\cmidrule(r){6-7}
		\cmidrule(r){8-10}
		& & & & & p & $\#$conf & $\#$conf & $\mathrm{m_{s}^2\chi_{disc}/T_{c}^{4}}$&  $\mathrm{m_{s}^2(\chi_{\pi} - \chi_{\delta})/T_{c}^{4}}$\\ \midrule \hline
		\multirow{8}*{6.664} &  \multirow{8}*{0.0514} &  0.002570  & 160 & $32^3\times8$  & 200000  & 2000 & 4054 &1.9(1) & 2.0(1)\\
		&  & 0.001904 & 140 & $32^3\times8$ & 200000 &  2000 &  6707 & 1.15(8) & 1.26(8)\\
		&  & 0.001285 & 110 & $40^3\times8$ & 200000 &  2000 &  5625 & 0.60(3) & 0.66(4)\\
		&  & 0.0006425 & 80 & $32^3\times8$ & 300000 &  1891 & 3842 & 0.14(1) & 0.18(2)\\
		&  & 0.0006425 & 80  & $40^3\times8$ & 300000  &  1998 & 11863 & 0.20(1)& 0.25(2)\\
		&  & 0.0006425 & 80  & $56^3\times8$   & 300000 &  1992 &  7341 & 0.20(1)& 0.24(2)\\
		&  & 0.0006425 & 80  & $72^3\times8$ & 300000 &  2088 & 5954 & 0.171(7)& 0.20(1)\\
		&  & 0.00032125 & 55  & $56^3\times8$ & 300000  &  2000 & 8473 & 0.059(7) & 0.08(1)\\ \midrule\hline
		\multirow{6}*{7.078} &   \multirow{6}*{0.034} &  0.00170  & 160 & $48^3\times12$  & 100000  & 3614  & 8507 & 5.8(5)& 5.9(4)\\
		&  & 0.001259 & 140 & $48^3\times12$  & 100000 & 2000 &  6575 & 4.1(2)& 4.3(2)\\
		&  & 0.000850  & 110 & $60^3\times12$ & 100000 &  2990  &  5314  & 2.0(2) & 2.1(2)\\
		&  & 0.000425 & 80 &  $48^3\times12$  & 300000 &  1993  & 9021 & 0.89(8)& 1.0(1)\\
		&  & 0.000425 & 80 &  $60^3\times12$  & 300000 &  1998 & 6746 & 0.78(6) & 0.91(7)\\
		&  & 0.000425 & 80 &  $72^3\times12$  & 200000 &  2365 &  2365 & 0.9(1) & 1.1(2)\\ \midrule \hline
		\multirow{5}*{7.356} &   \multirow{5}*{0.026}  &  0.0013 & 160 & $64^3\times16$  & 100000 & 2198 & 3227  &12(1) & 12(1)\\
		&  & 0.000963 & 140 & $64^3\times16$  & 100000 & 2370 &  3639 & 11(2) & 10(2)\\
		&  & 0.000650  & 110 & $64^3\times16$ & 100000  &  2321 &  3498 & 7(1) & 7(1)\\
		&  & 0.000325 & 80 & $64^3\times16$ & 500000 &  2577 &  4092 & 3.4(4)& 3.7(4)\\
		&  & 0.000325 & 80 & $80^3\times16$ & 300000 &  3001 & 5316 & 3.0(2)& 3.7(3)\\ \midrule
		\bottomrule
	\end{tabular}
	\caption{Summary of lattice parameters, i.e. values of lattice gauge coupling $\beta$, strange ($am_s$) and light quark mass  ($am_l$) in unit of lattice spacing, pion mass ($m_\pi$), lattice size ($N_{\sigma}^3\times N_{\tau}$), number of gauge configurations ($\#$conf) used in the direct measurements of chiral observables (Direct) and in the computation of $\rho$ via the Chebyshev polynomial method (Chebyshev) as well as the order of Chebyshev polynomials ($p$). The direct measurements of $\mathrm{m_{s}^2\cdsc/T_{c}^{4}}$ and $\mathrm{m_{s}^2(\chi_{\pi} - \chi_{\delta})/T_{c}^{4}}$ with $T_c=132$ MeV are also listed.}
	\label{sup:table_setup}
\end{table}

	\begin{figure*}[!thp]
		\includegraphics[width=0.43\textwidth]{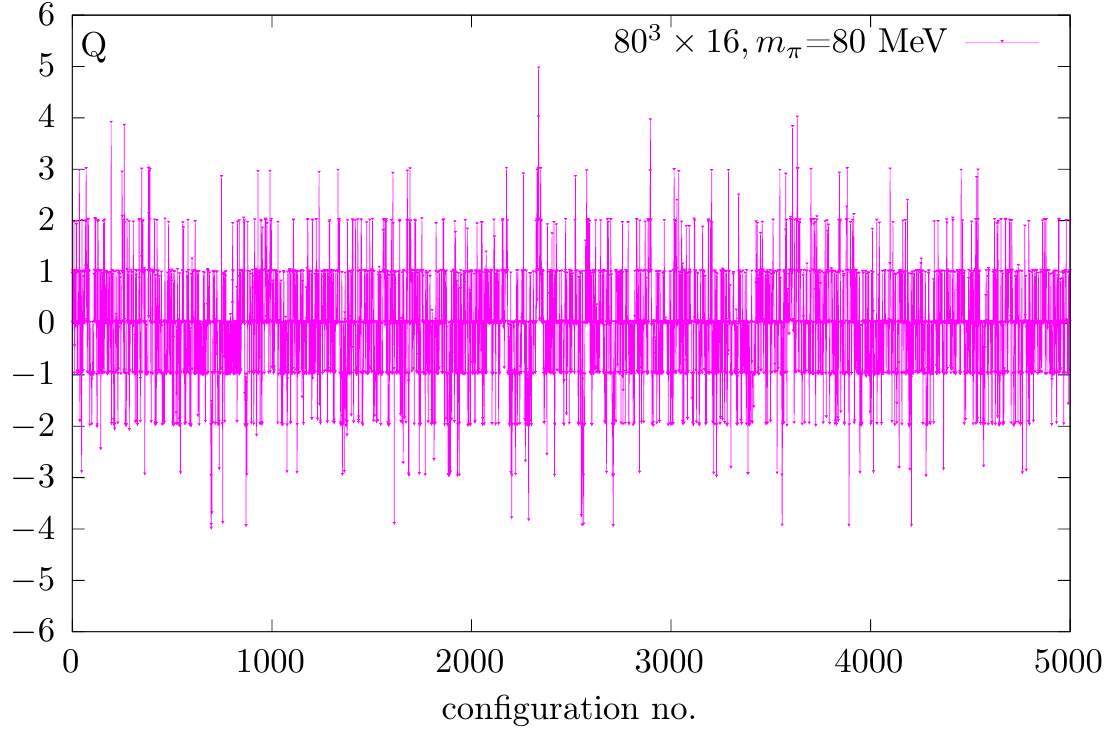}
		\includegraphics[width=0.43\textwidth]{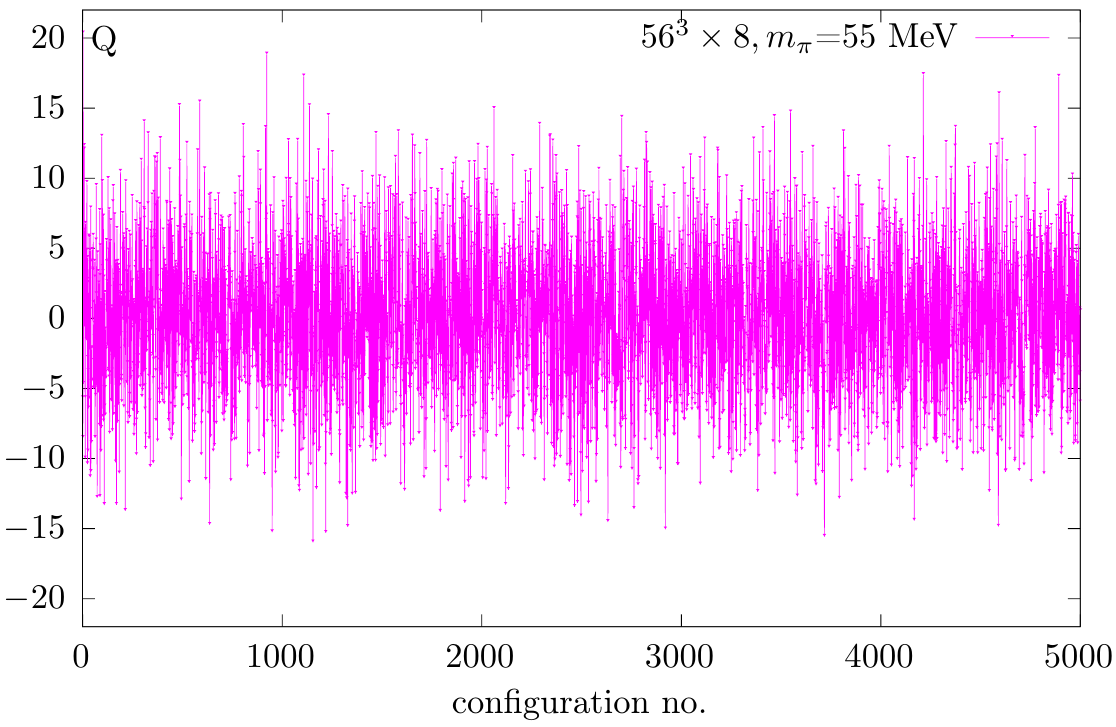}
		\caption{Left: Time history of topological charge obtained from $80^3\times16$ lattices with $m_\pi=80$ MeV (left) and $56^3\times8$ lattices with $m_\pi=55$ MeV (right) using the Symanzik improved gradient method~\cite{Ramos:2015baa}. Results shown in both plots are obtained at a certain flow time $t$.
			The value of $t$ is chosen to be $T\sqrt{8t}=0.4$ where the topological charge susceptibility already reaches to a plateau in $t$. This corresponds to $t/a^{2}=0.02N_{\tau}^{2}$. }
		\label{fig:sup_top_history}
	\end{figure*}

	\subsection{II B. Methodology on the computation of $\rho$}
In this subsection we describe the method we used to compute the spectral density of the lattice Dirac operator as a continuous function over all scales of the complete eigenvalue spectrum. The method has been utilized in the Wilson~\cite{Giusti:2008vb,Patella:2012da}, Domain Wall~\cite{Cossu:2016eqs} and staggered~\cite{Fodor:2016hke, Ding:2020hxw,Ding:2020eql,Itou:2014ota,deForcrand:2017cja} discretization schemes.  In the following subsection we will also present sanity checks of this method.

Stochastic counting of eigenvalues of a hermitian matrix $A$ in a given interval $[s,t]$ within $[-1,1]$ can be represented as
\begin{eqnarray}
n[s,t] & = &\frac{1}{N_r} \sum_{r=1}^{N_r}\xi_{r}^{\dagger}h(A)\xi_r\, ,
\label{eq:sup_mn}
\end{eqnarray}
where ${N_r}$ is the number of random vectors, $\xi_{r}$ is a Gaussian random noise vector and $h(A)$ is a step function which equals to 1 only in the interval $[s,t]$ and 0 elsewhere. Here $t>s$ is implicitly assumed. In practice, the function $h(A)$ is approximated by the Chebyshev polynomial
\begin{eqnarray}
h(A) &=& \sum_{j=0}^{p}g_j^{p}\gamma_{j}T_{j}(A). \label{eq:sup_hA}
\end{eqnarray}
The coefficients $g_j^{p}$ and $\gamma_{j}$ are known numbers once the interval $[s,t]$ is given, and $p$ is the order of Chebyshev polynomials. As the expansion of $h(A)$ has harmful oscillations near the boundaries $g_j^{p}$ is introduced \cite{NapoliPS13} here to suppress this behavior. $T_{j}(A)$ is the Chebyshev polynomial of operator $A$ and it can be constructed by the following recursion relation
\begin{equation}
T_0(A) = 1,\quad T_1(A) = A,\quad T_j(A) = 2AT_{j-1}(A) - T_{j-2}(A) \quad(j\geq 2 ).
\label{eq.test3}
\end{equation}

\begin{figure*}[!thp]
	\includegraphics[width=0.45\textwidth]{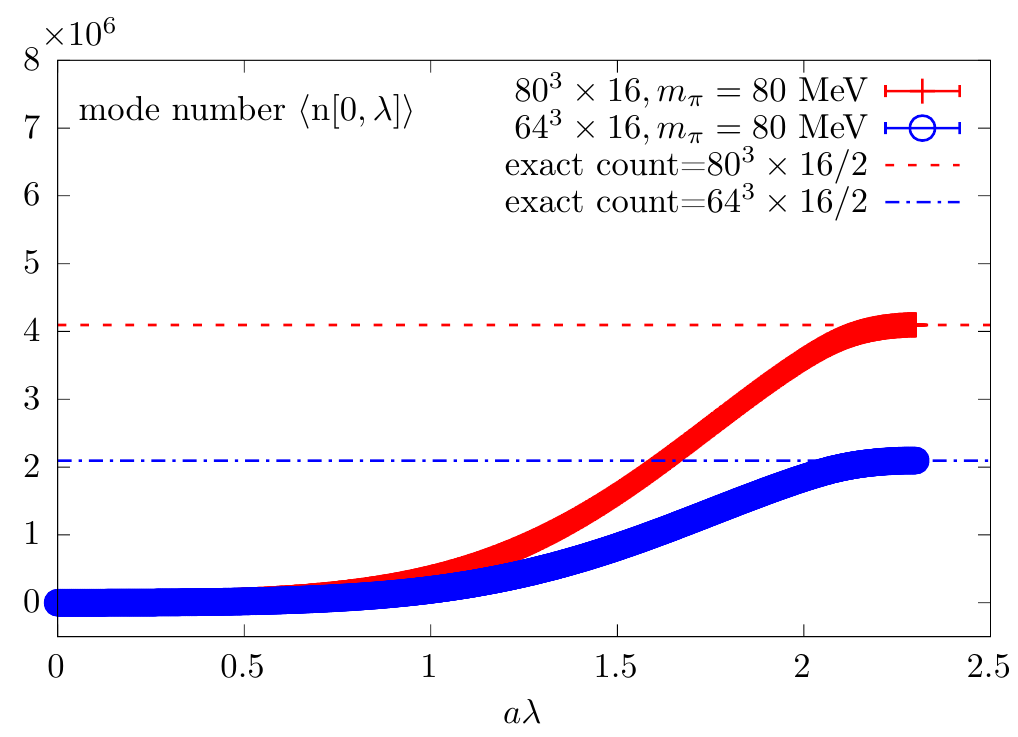}~~
	\caption{Averaged mode number $n[0,\lambda]$ over configurations as a function of $\lambda$ in unit of lattice spacing $a$ computed on $80^3\times16$ and $64^3\times16$ lattices with $m_\pi=80$ MeV. The mode numbers obtained from $80^3\times16$ and $64^3\times16$ lattices reach to the numbers of $80^3\times16/2$ and $64^3\times16/2$, respectively, in the large $\lambda$ limit for positive $\lambda$ as expected.}
	\label{fig:sup_mn}
\end{figure*}

The above deviation is based on the assumption that all the eigenvalues of $A$ are restricted in the range of $[-1,1]$. In order to apply the eigenvalue filtering method to calculate the Dirac spectrum, we therefore define
\begin{equation} 
A = \frac{\slashed{D}_{stag}^{\dagger}\slashed{D}_{stag} - \frac{(\tilde{\lambda}_{{\rm max}} + \tilde{\lambda}_{{\rm min}})}{2} \mathbbm{1} }{\frac{(\tilde{\lambda}_{{\rm max}} -\tilde{\lambda}_{{\rm min}})}{2} \mathbbm{1}},
\label{eq:sup_Alambda}
\end{equation}
such that the eigenvalues of $A$ are all distributed in $[-1,1]$. Here $\slashed{D}_{stag}$ stands for the massless Dirac matrix in the staggered discretization scheme, i.e. $\slashed{D}_{stag}$ is defined as $M_{stag}$ in the case of vanishing quark mass.
Substituting the expression of $h(A)$ (\autoref{eq:sup_hA}) into the stochastic estimator (\autoref{eq:sup_mn})  we can obtain the mode number ${n}[s,t]$ for a given gauge configuration,
\begin{equation}
{n}[s,t] = \frac{1}{N_r} \sum_{r=1}^{N_r}  \sum_{j=0}^p g^{p}_j \gamma _{j} \,{ \xi_r^\dagger T_{j}(A) \xi_r }.
\label{eq.test5}
\end{equation}

Once we get the mode number $n[s,t]$, $\rho_U(\lambda)=\sum_j\delta(\lambda-\lambda_j)$ can be easily constructed as
\begin{equation}
\rho_U(\lambda) =  \frac{1}{4} \frac{n[s,t]}{2\,\delta\lambda},
\label{eq.test6}
\end{equation}
where the factor $1/4$ accounts for the fourth-root arising from the staggered discretization scheme, 
the factor 2 in the denominator is due to the positive and negative eigenvalue pairs, $\delta\lambda$ is the bin-size,  and $\lambda\equiv\lambda^{\vert i\slashed{D}_{stag} \vert}$ is related to $\tilde\lambda$ being the eigenvalues of matrix $\slashed{D}^\dagger_{stag}\slashed{D}_{stag}$ as follows
\begin{align}
\begin{split}
\lambda &= \sqrt{\tilde{\lambda}} = \left[s\,\left(\tilde{\lambda}_{\rm max} - \tilde{\lambda}_{\rm min}\right)/2 +  \left(\tilde{\lambda}_{\rm max} + \tilde{\lambda}_{\rm min}\right)/2 \right]^{1/2},\\
\lambda +\delta\lambda &= \sqrt{\tilde{\lambda}} +\delta\lambda= \left[t\,\left(\tilde{\lambda}_{\rm max} - \tilde{\lambda}_{\rm min}\right)/2 +  \left(\tilde{\lambda}_{\rm max} + \tilde{\lambda}_{\rm min}\right)/2 \right]^{1/2}.
\end{split}
\end{align}
In our work $\tilde{\lambda}_{{\rm min}}$ being the minimum value 
of $\tilde{\lambda}$ is set to 0 and $\tilde{\lambda}_{{\rm max}}$ 
being the maximum value of $\tilde{\lambda}$ is estimated 
by the power method~\cite{PowerMethod:Saad92}.  The ensemble 
averaged $n[0,\lambda]$ obtained from two different volumes of 
lattices is demonstrated in~\autoref{fig:sup_mn}.

Once $\rho_U$ is obtained it is straightforward to compute the Dirac eigenvalue 
spectrum $\rho$ as $T/V$ multiplied by the average of $\rho_U$ 
over gauge configurations, i.e. $\rho\equiv\frac{T}{V}\langle\rho_U\rangle$ (\cf~\autoref{eq:ro}).
Similarly one can also compute the correlation functions as $C_n(\lda_1, \cdots, \lda_n; \ml) =
\av{\prod_{i=1}^n \qty[ \ru{i} - \av{\ru{i}} ] }$. The error analyses of $\rho$ and $C_n$ presented in our paper are all done using the Jackknife method.  

For demonstration we show $\rho$ and $C_2(\lambda_1,\lambda_2)$ obtained from $80^3\times16$ lattices with $m_\pi=$80 MeV in the left and right panel of \autoref{fig:sup_C2}, respectively. 
\begin{figure*}[!thp]
	\includegraphics[width=0.45\textwidth,height=0.25\textheight]{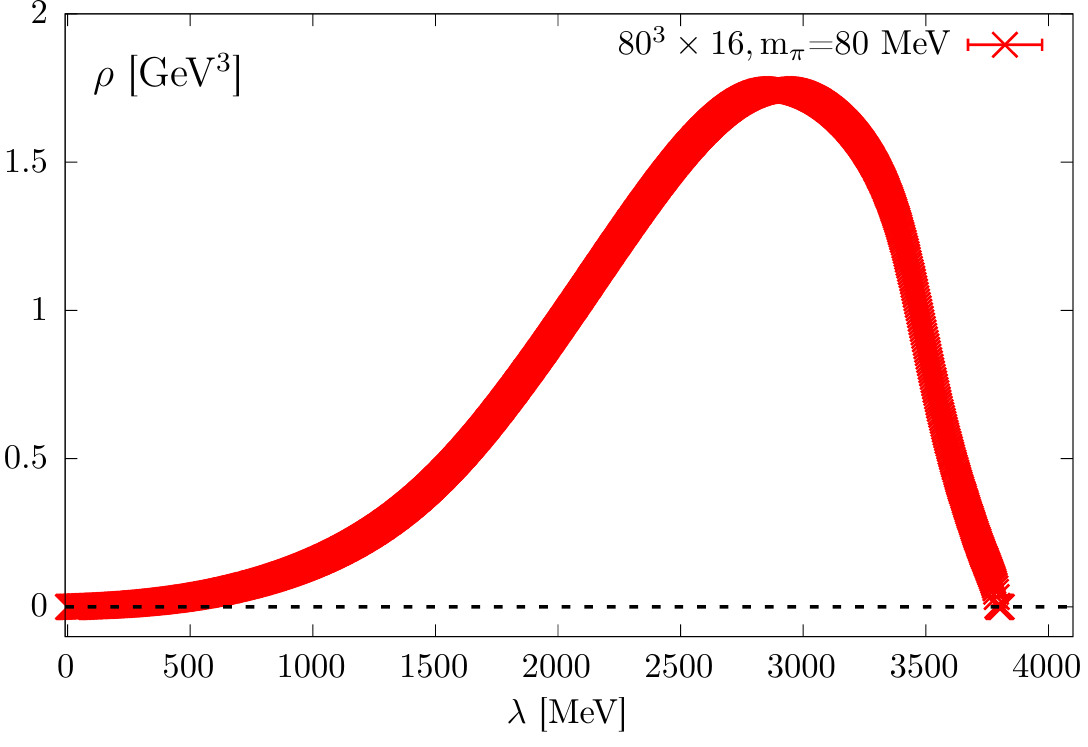}~
	\includegraphics[width=0.42\textwidth]{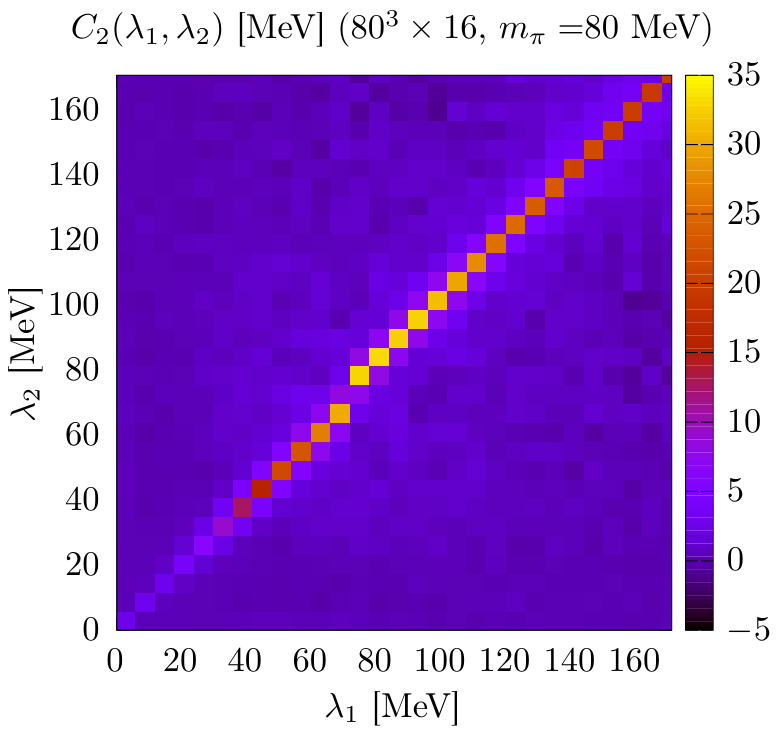}
	\caption{Dirac eigenvalue spectrum $\rho$ (left) and the two-point correlation function $C_2$ (right) obtained on $80^3\times16$ lattices with $m_\pi=$80 MeV.}
	\label{fig:sup_C2}
\end{figure*}

\subsection{II C. Sanity checks of $\pdv*[n]{\rho}{m_l}$}
			\begin{figure*}[!thp]
	\includegraphics[width=0.4\textwidth]{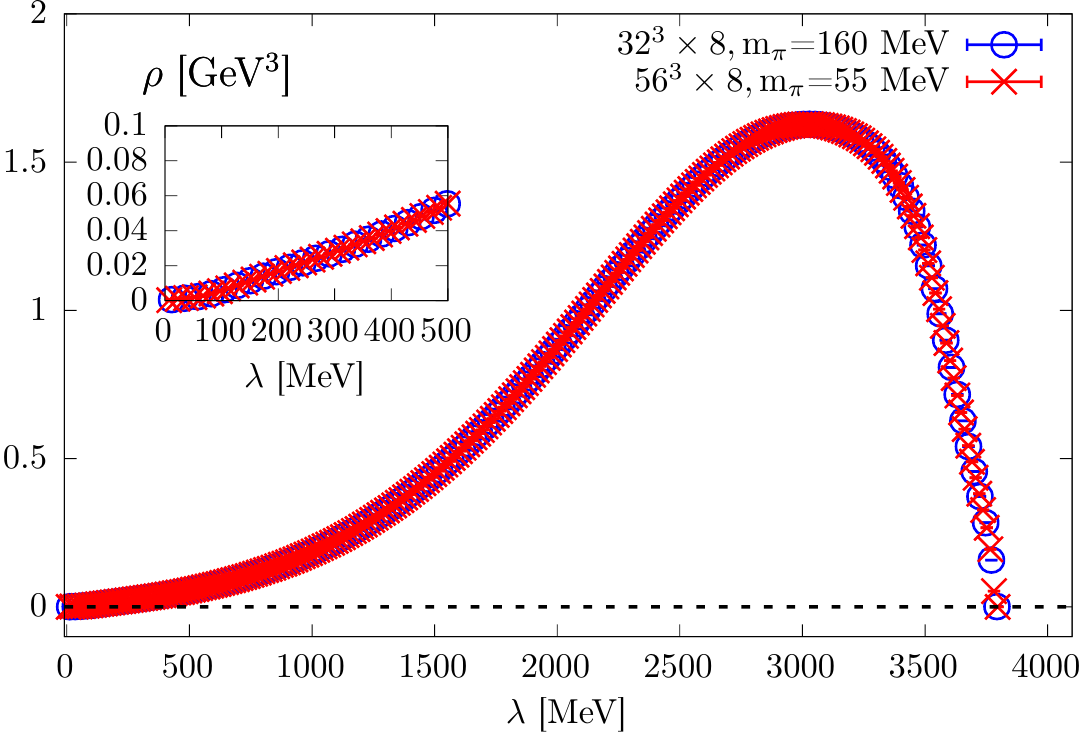}~
	\includegraphics[width=0.4\textwidth]{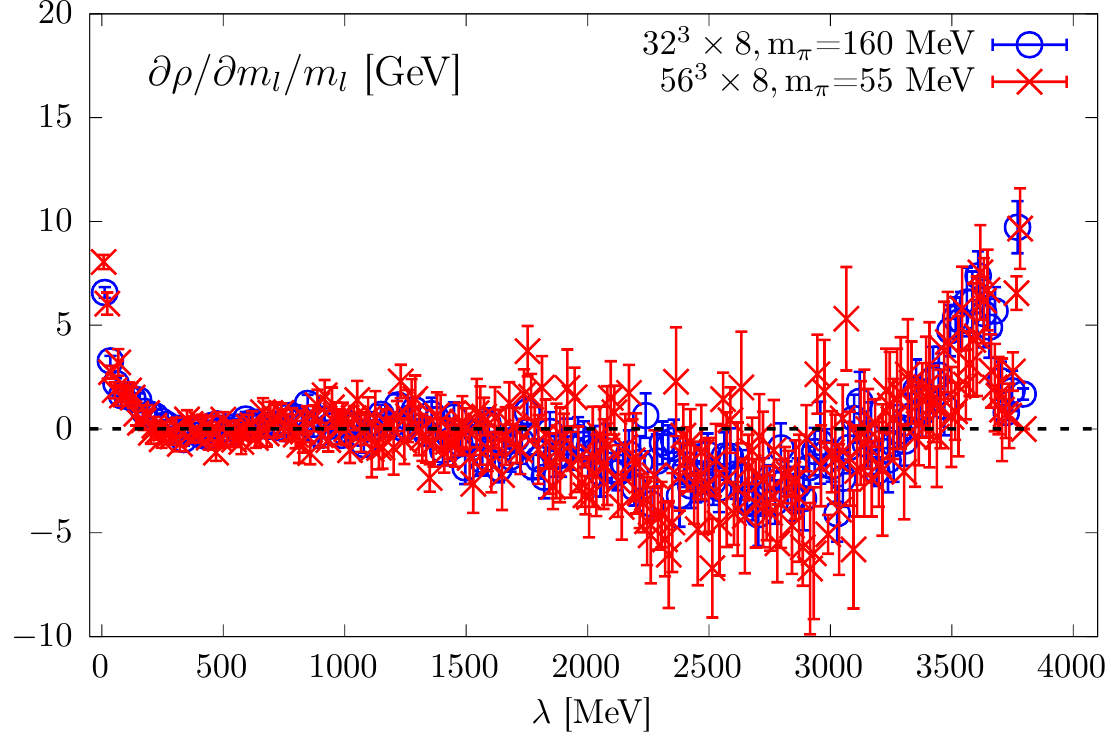} ~
	\includegraphics[width=0.4\textwidth]{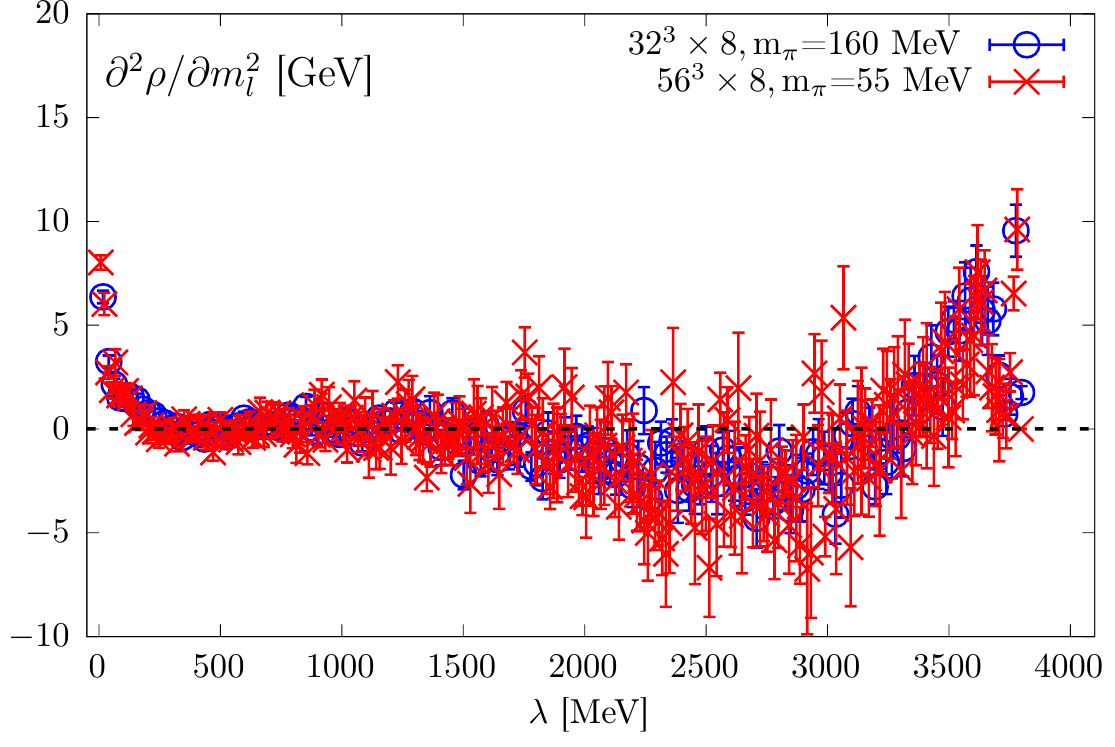}~	
	\includegraphics[width=0.4\textwidth]{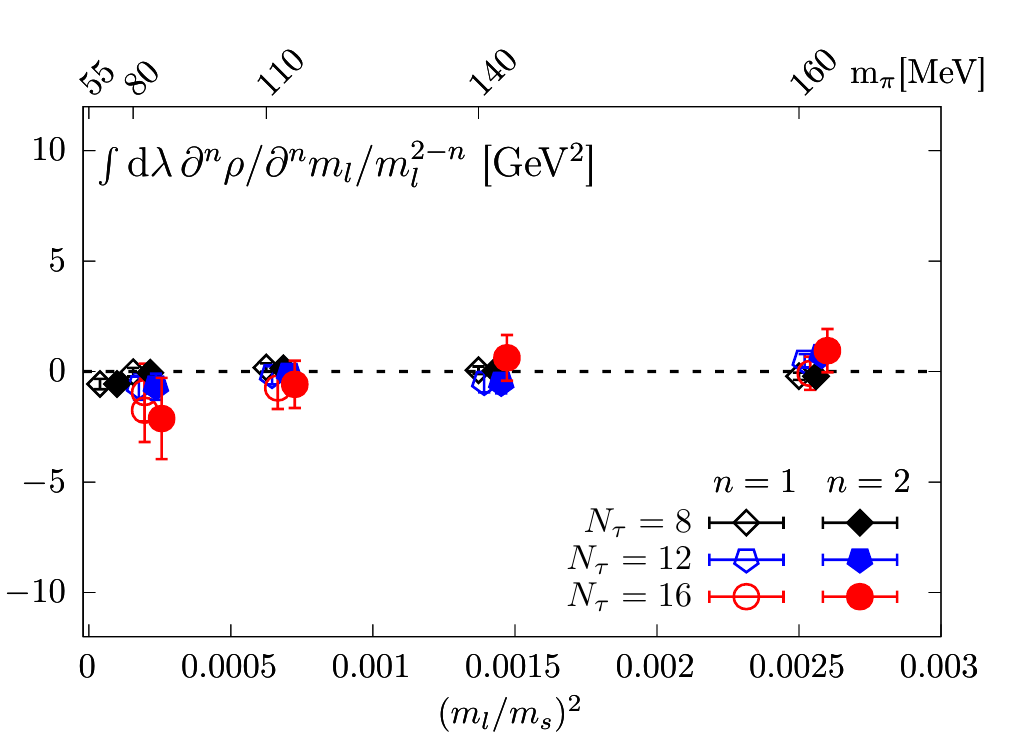}
	\caption{$\rho$ (top left), $m_l^{-1}\pdv*[]{\ro}{\ml}$ (top right) and $\pdv*[2]{\ro}{\ml}$ (bottom left) as a function of $\lambda$ in a complete region of $\lambda$. The insert in the top left panel shows a blow-up of $\rho$ in the region of $\lambda\in[0,500]$ MeV.  In these three plots results are all obtained from $N_\tau=8$ lattices with $m_\pi=160$ and 55 MeV. The bottom right panel shows the integration of $\int \mathrm{d}\lambda\,m_l^{n-2}\pdv*[n]{\ro}{\ml}$ for $n=1$ and 2 obtained at all the quark masses and lattice spacings with the largest $N_\sigma$ available. The filled symbols are slightly shifted horizontally for visibility.}
	\label{fig:sup_complete_rho}
\end{figure*}

As from the definition of $C_n$ the following constraint is fulfilled
\begin{equation}
\int_0^\infty d {\lambda_i} \,C_n(\lambda_1,\lambda_2,\cdots,\lambda_i,\cdots,\lambda_n;m_l) = 0\,,
\end{equation}
where $\lambda_i$ stands for any one of $\lambda$'s. Consequently 
\begin{equation}
\int_0^\infty d {\lambda} \,\pdv*[n]{\ro}{\ml}= 0\,,\quad \quad \mathrm{with} \quad n\geq 1~\mathrm{and} ~n\in \mathbb{Z}\,.
\label{eq:sup_dnrho_constraint}
\end{equation}
\autoref{eq:sup_dnrho_constraint} thus suggests that $\pdv*[n]{\ro}{\ml}$ with $n\geq1$ must either contain both negative and positive parts in $\lambda$ or vanish for all values of $\lambda$.

Here we demonstrate the complete spectrum of $\rho$ and its first and second derivatives in $m_l$ obtained from $N_\tau=8$ lattices with $m_\pi=160$ and 55 MeV. We first show the complete spectrum of $\rho$ in the top left panel of~\autoref{fig:sup_complete_rho}. It can be found that the $m_l$ dependence can be hardly observed from $\rho$ directly. In the top right and bottom left panels of ~\autoref{fig:sup_complete_rho} we show
complete spectrum of $m_l^{-1}\pdv*[]{\ro}{\ml}$ and $\pdv*[2]{\ro}{\ml}$, respectively. It can be clearly seen that both $m_l^{-1}\pdv*[]{\ro}{\ml}$ and $\pdv*[2]{\ro}{\ml}$ possess negative and positive values in the complete $\lambda$ region. In the bottom right panel of~\autoref{fig:sup_complete_rho} we show the integrations $\int\mathrm{d}\lambda\,m_l^{n-2}\pdv*[n]{\ro}{\ml}$ for $n=1$ and 2 obtained at all the quark masses and lattice spacings with the largest $N_\sigma$ available. It can be observed that all these integrations are consistent with zero within errors as expected ($\cf$~\autoref{eq:sup_dnrho_constraint}).

\begin{figure*}[!htp]
	\includegraphics[width=0.44\textwidth]{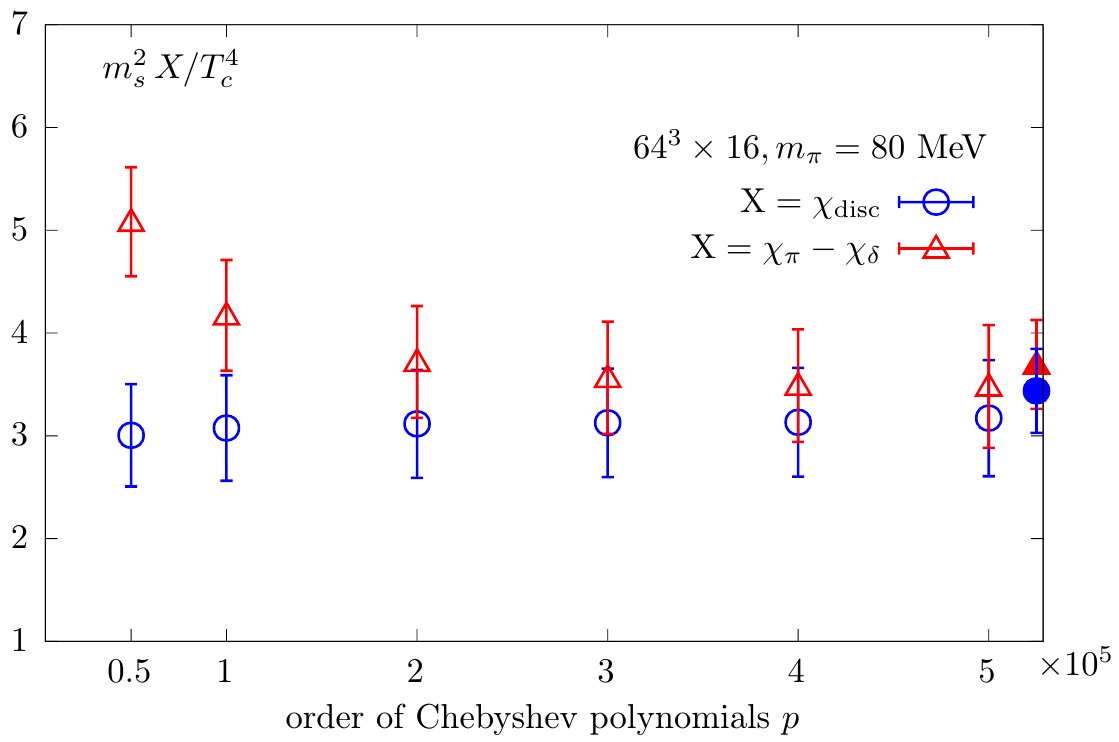}
	\caption{Dependence of $\cdsc$ and $\chi_\pi-\chi_\delta$ obtained from $\rho$ and $\partial\rho/\partial m_l$ on the order of Chebyshev polynomials $p$. The corresponding filled symbols located on the most right hand side denote the results obtained from direct measurements. The computation were performed on $64^3\times16$ lattices with $m_\pi=80$ MeV.}
	\label{fig:sup_chi_pdep}
\end{figure*}	
The final results of $\rho$ depend on the order of Chebyshev polynomials $p$. We show in~\autoref{fig:sup_chi_pdep} the $p$-dependence of $\chi_\pi-\chi_\delta$ and $\chi_{disc}$ 
computed via
~\autoref{eq:sus-rho1} and~\autoref{eq:sus-rho2} 
, respectively, from $64^3\times16$ lattices with $m_\pi=80$ MeV. It can be seen that with $p\geq100000$
	results of both $\chi_\pi-\chi_\delta$ and $\chi_{disc}$ start to saturate and agree within errors with the results obtained from direct measurements.

\section{III. Results}
\subsection{III A. Supplemental materials to~\autoref{fig:rho}}
We show similar results to the left panel of~\autoref{fig:rho}
 but for $N_t=12$ and 16 in the left and right panels 
of~\autoref{fig:sup_fig1left}, respectively. The
 general feature observed in the left panel of~\autoref{fig:rho} persists in the results obtained from finer lattices.
\begin{figure*}[!htp]
	\centering
	\includegraphics[width=0.4\textwidth]{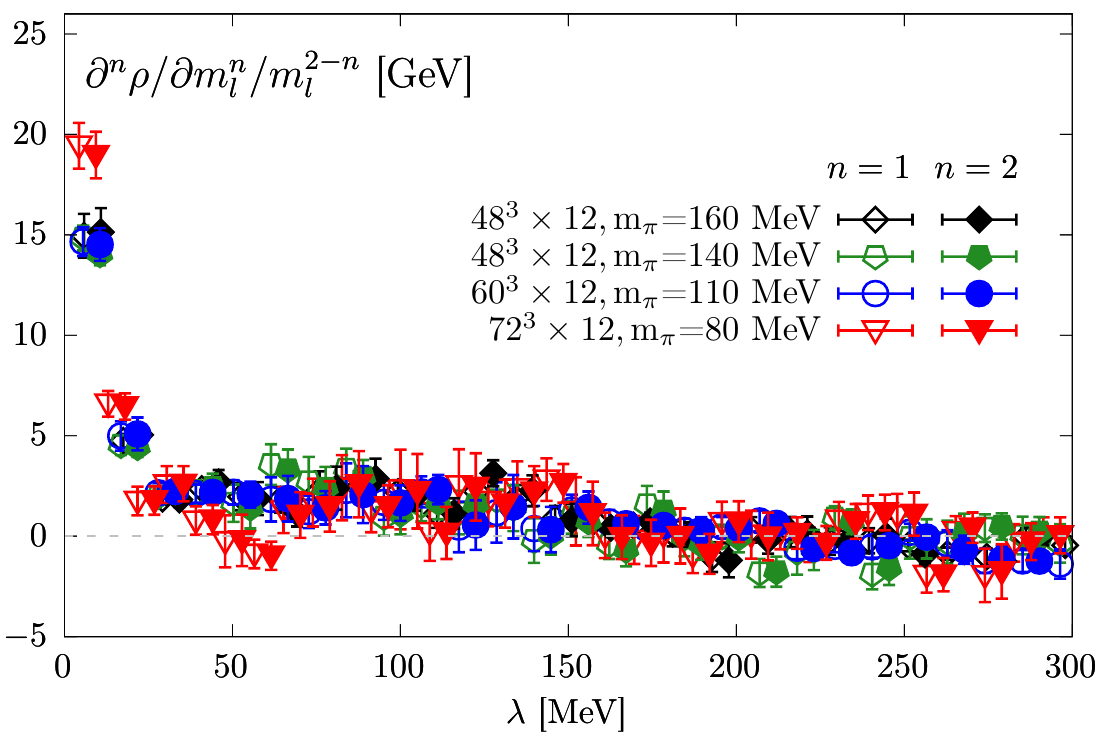}~~
		\includegraphics[width=0.4\textwidth]{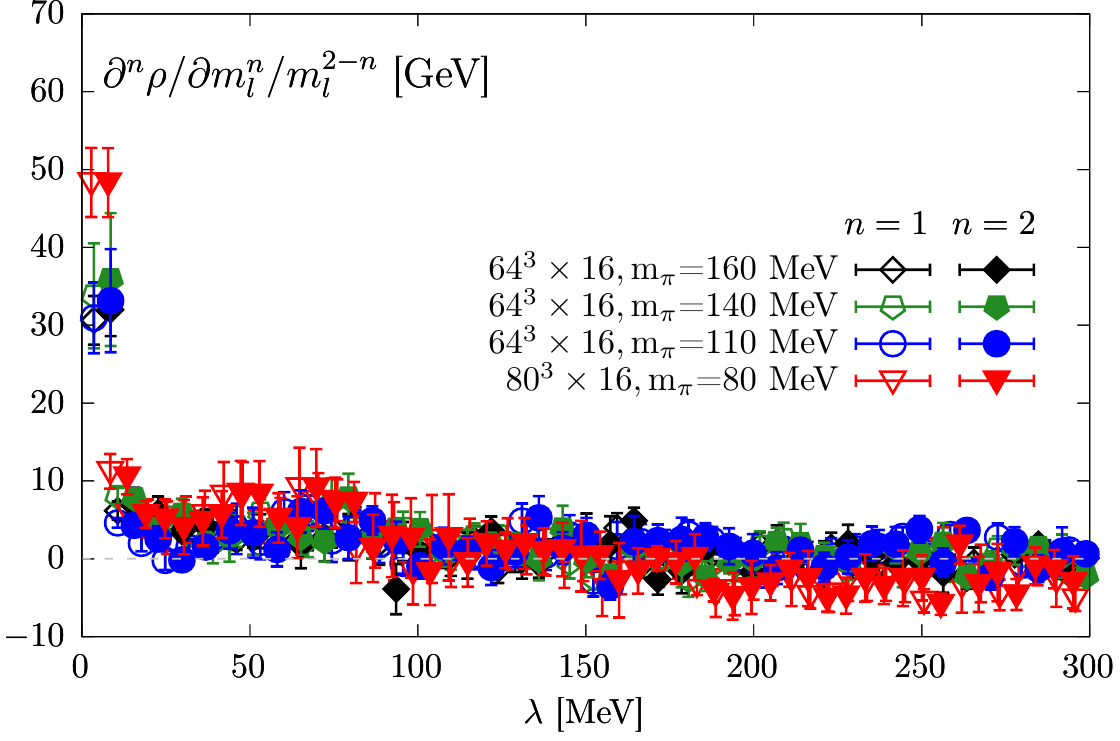}
	\caption{Light sea quark mass dependence of $m_l^{-1}\pdv*{\ro}{\ml}$ (open
		symbols) and $\pdv*[2]{\ro}{\ml}$ (filled symbols) using $\nt=12$ (left) and $\nt=16$ (right) lattices. In all cases,
		results are obtained at $T\approx205$~MeV and the filled symbols have been slightly
		shifted horizontally for visibility.
	}
\label{fig:sup_fig1left}
\end{figure*}

Supplementary to the results shown in the middle panel of~\autoref{fig:rho} we show results of $\pdv[2]{\rho}{\ml}$ and $\pdv[3]{\rho}{\ml}$ (inset) at $m_\pi$=110 MeV (left), 140 MeV (middle) and 160 MeV (right) in~\autoref{fig:sup_fig1middle1}. 
\begin{figure*}[!htp]
	\centering
	\includegraphics[width=0.32\textwidth]{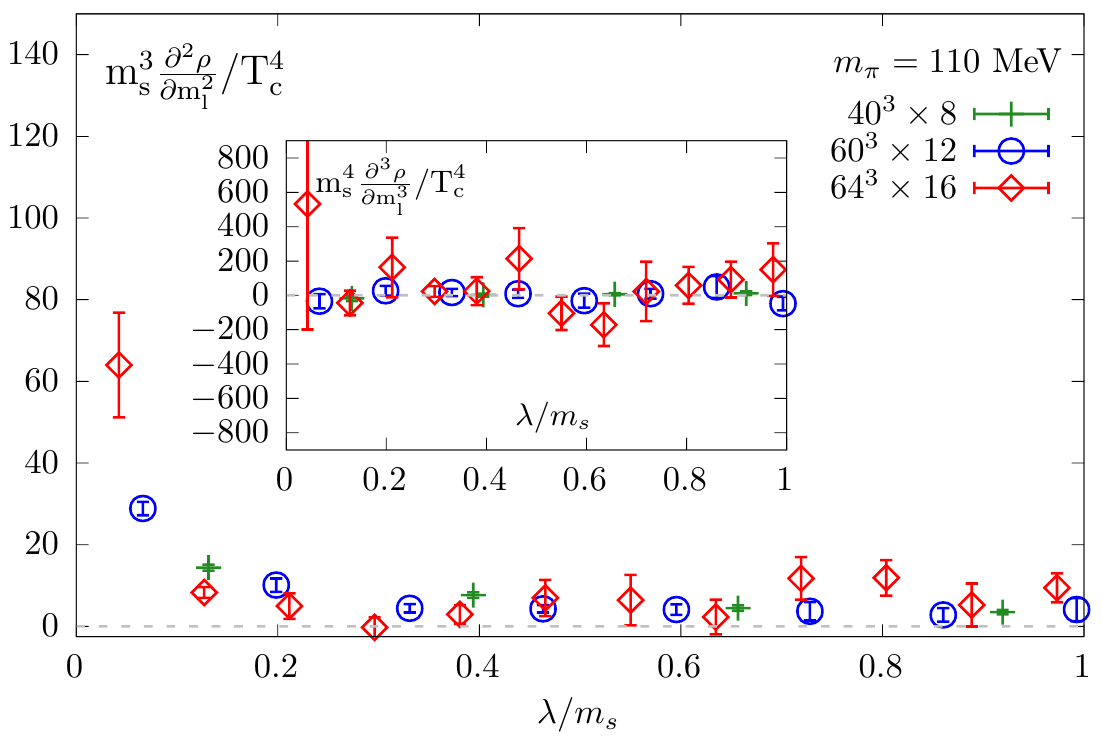}~
	 \includegraphics[width=0.32\textwidth]{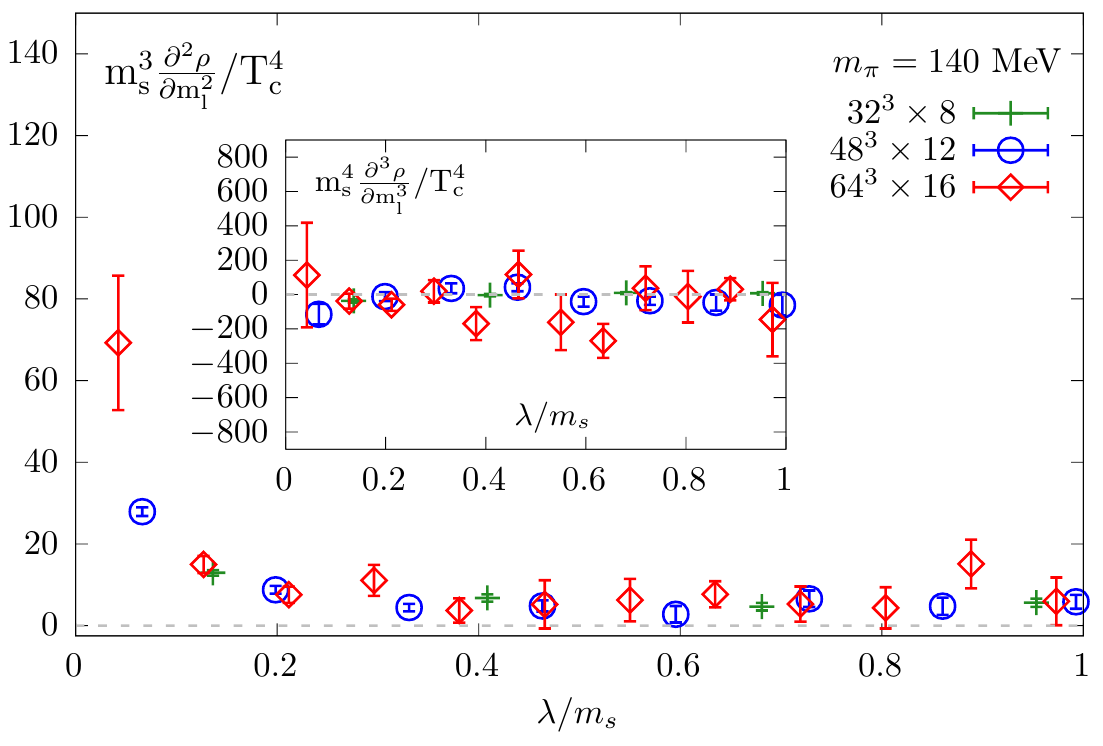}~
 	\includegraphics[width=0.32\textwidth]{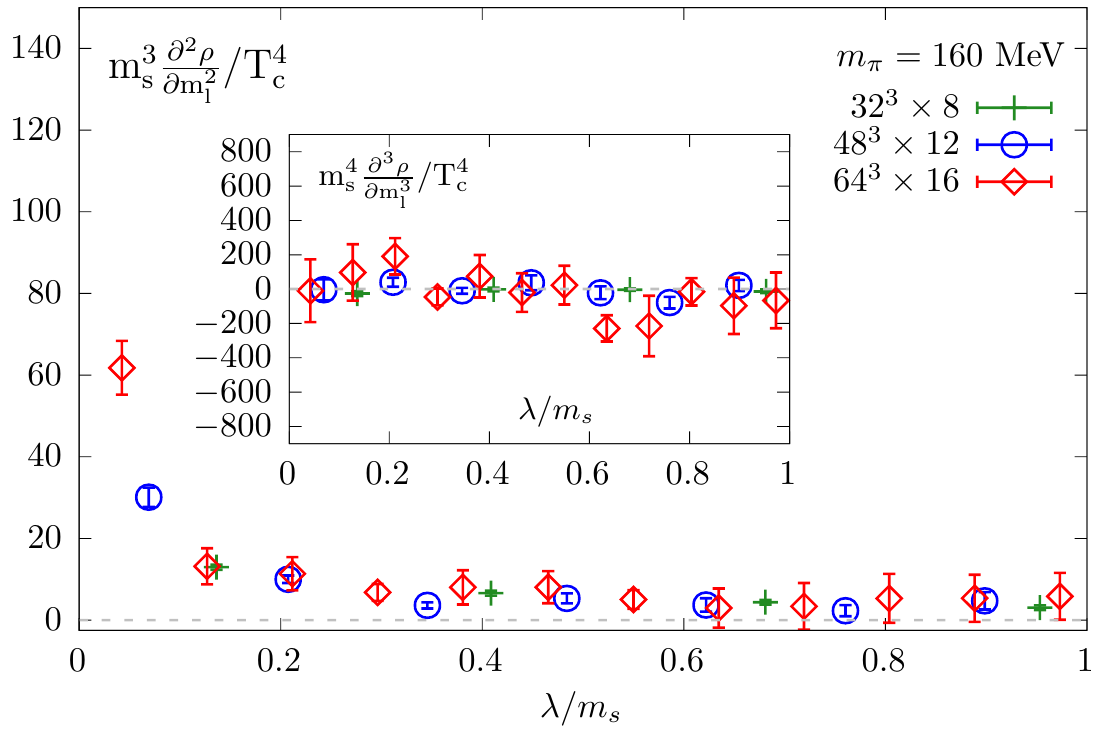}
	\caption{Lattice spacing dependence of $\pdv*[2]{\ro}{\ml}$ and
		$\pdv*[3]{\ro}{\ml}$ (inset) for $m_\pi=110$~MeV (left), 140 MeV (middle) and 160 MeV (right). In all cases,
		results are obtained at $T\approx205$~MeV.}
	\label{fig:sup_fig1middle1}
\end{figure*}

\begin{figure*}[!htp]
	\centering
	\includegraphics[width=0.6\textwidth]{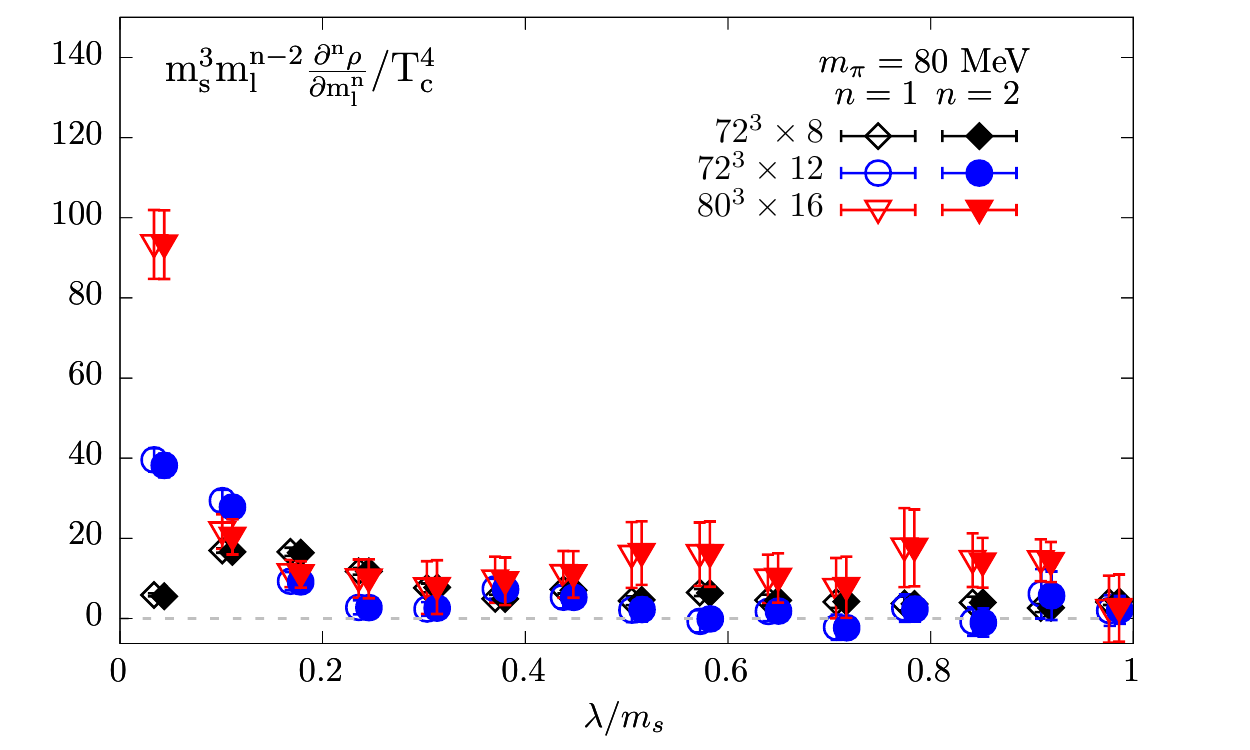}~
	\caption{Lattice spacing dependence of $m_l^{-1}\pdv*{\ro}{\ml}$ (open
		symbols) and $\pdv*[2]{\ro}{\ml}$ (filled symbols) for $m_\pi=80$~MeV at the same bin-size in $\lambda/m_s$ as used on $N_\tau=16$ lattices in the main text. The filled symbols have been slightly
		shifted horizontally for visibility.}
		\label{fig:sup_fig1middle2}
\end{figure*}

We show in~\autoref{fig:sup_fig1middle2} the 
	lattice spacing dependence of $m_l^{-1}\pdv*{\ro}{\ml}$ and $\pdv*[2]{\ro}{\ml}$ for $m_\pi=80$~MeV at the
	same bin-size in $\lambda/m_s$ as used on $N_\tau=16$ lattices in the main text. It can be seen that the feature of a sharper peak persists towards the continuum limit.  It is expected that using this same bin-size in $\lambda/m_s$ the directly measured chiral observables for $N_\tau=12$ and $N_\tau=8$ lattices cannot be reproduced.

Supplementary to the results shown in the right panel of~\autoref{fig:rho} we show similar results but for $m_\pi$=110 MeV (left), 140 MeV (middle) and 160 MeV (right) in~\autoref{fig:sup_fig1right}.

\begin{figure*}[!htp]
	\centering
 \includegraphics[width=0.32\textwidth,height=0.17\textheight]{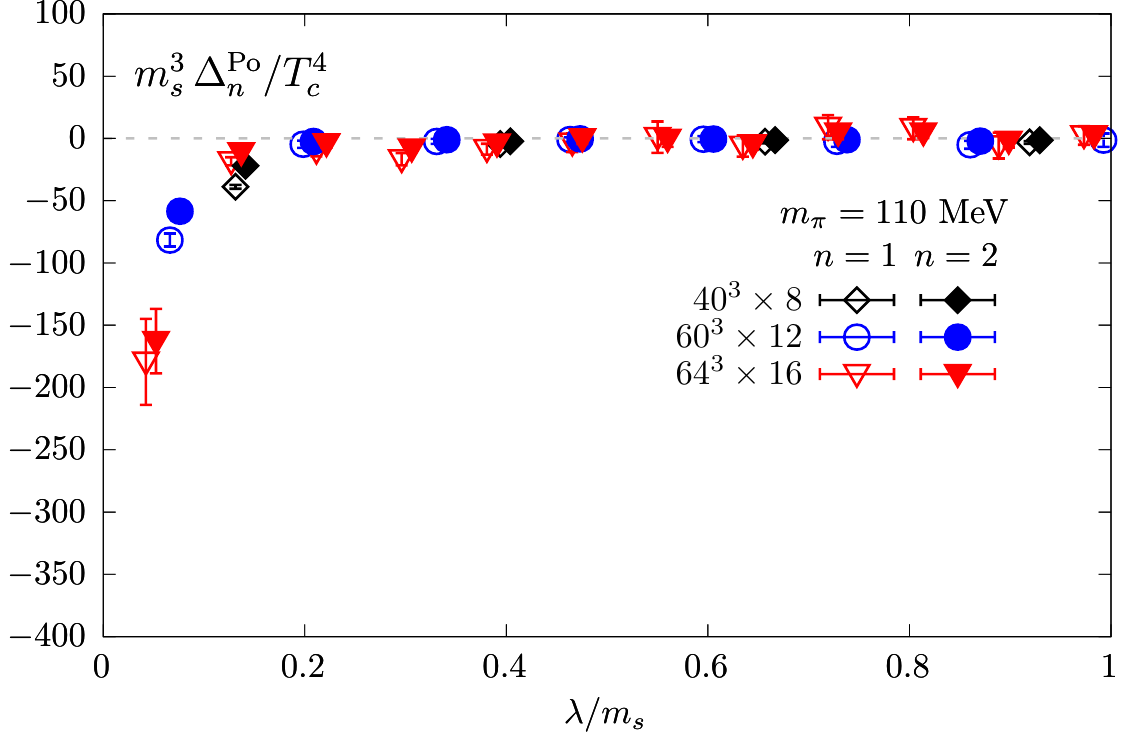}~
  \includegraphics[width=0.32\textwidth,height=0.17\textheight]{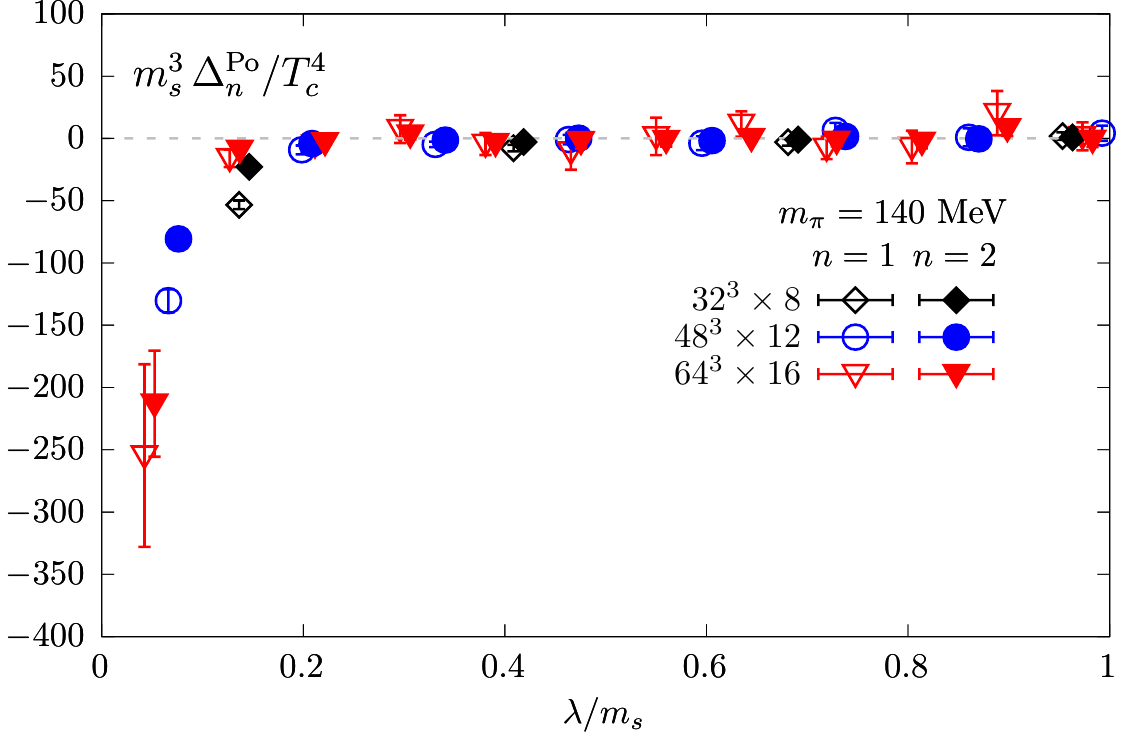}~
    \includegraphics[width=0.32\textwidth,height=0.17\textheight]{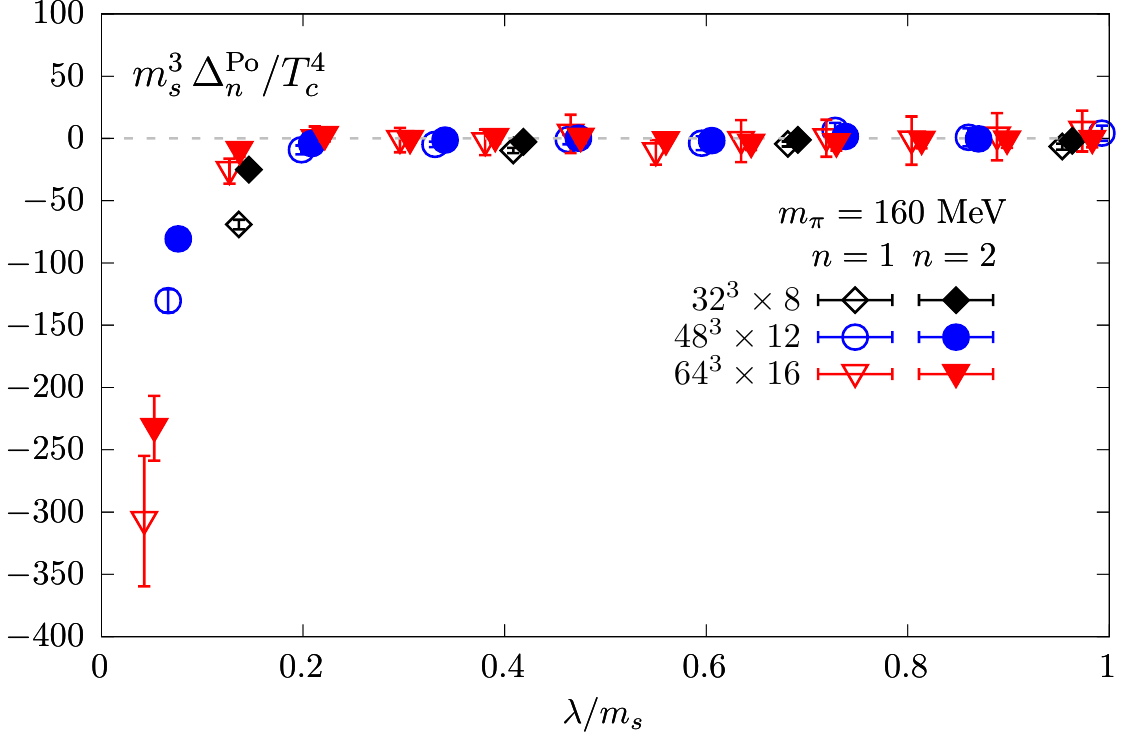}~
	\caption{The differences $\Delta_n^\mathrm{Po} = \ml^{n-2} \qty[ \pdv*[n]{\rho}{\ml} -
		(\pdv*[n]{\rho}{\ml})^\mathrm{Po} ]$ (\textit{cf.}~\autoref{eq:droPo}  and~\autoref{eq:dro2Po}), for $m_\pi=110$~MeV (left), 140 MeV (middle) and 160 MeV (right).}
		\label{fig:sup_fig1right}
\end{figure*}

\subsection{III B. Supplemental materials to~\autoref{fig:Comparison}}

	\subsubsection{III B1. Volume dependence of the two $U(1)_A$ measures}
	We show the volume dependence of  $\chi_\pi-\chi_\delta$ and $\chi_{disc}$ at $m_\pi=80$ MeV in the left and right panels of~\autoref{fig:sup_chidisc_V}, respectively. One can observe that the volume dependences of these two quantities are mild.
\begin{figure*}[!thp]
	\includegraphics[width=0.44\textwidth]{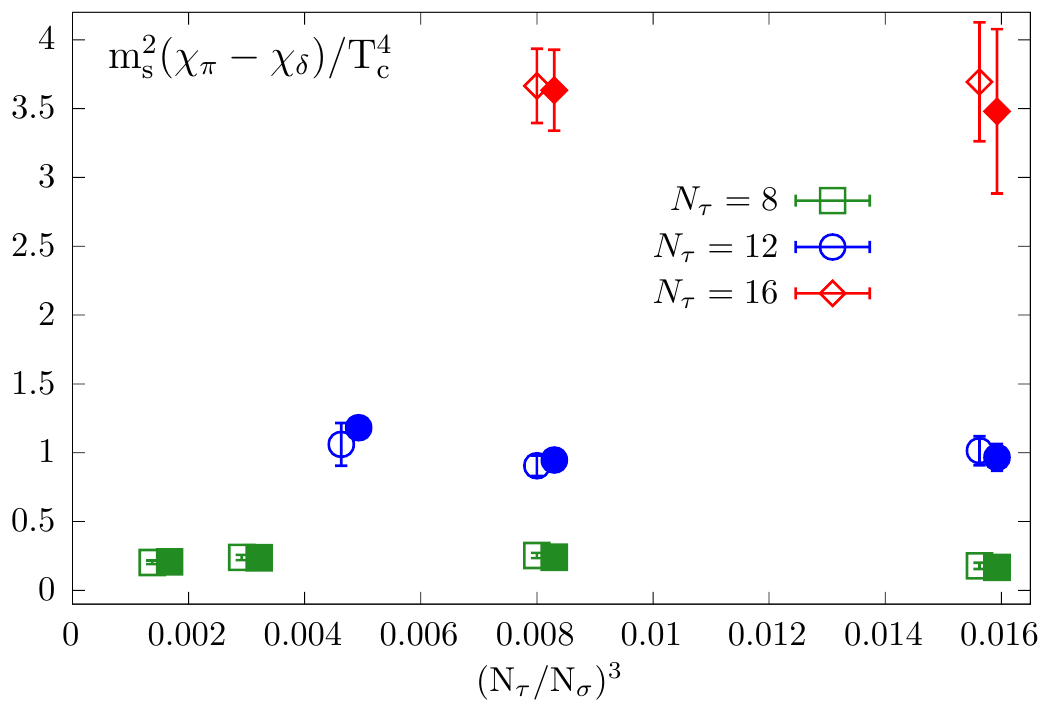}
	\includegraphics[width=0.44\textwidth]{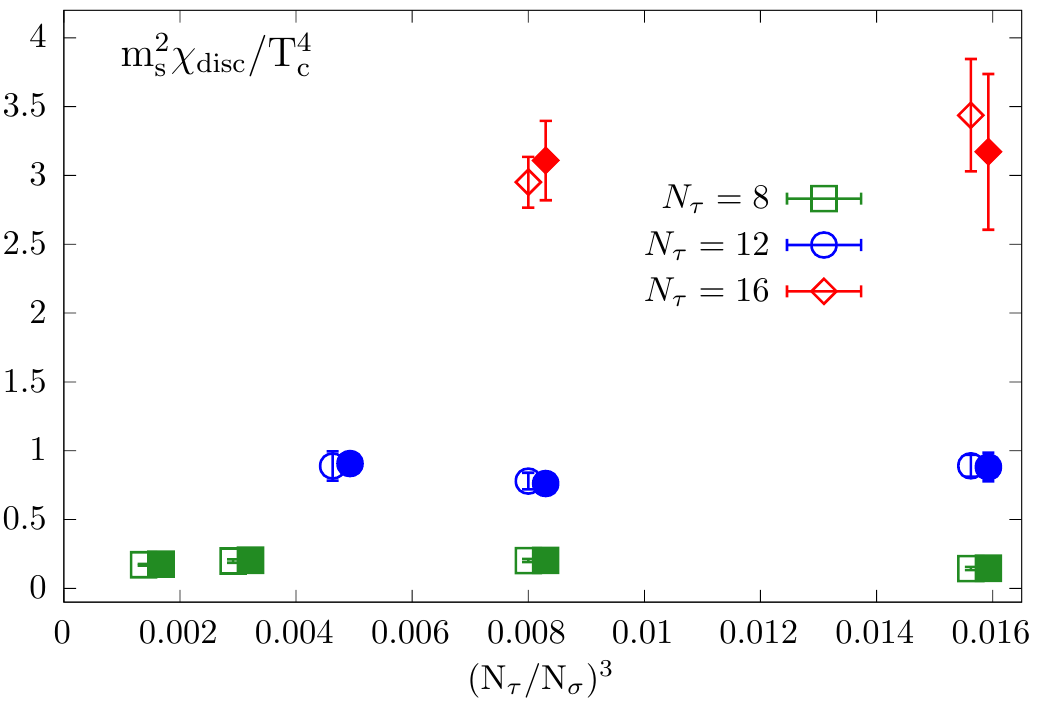}
	\caption{Volume dependences of $\chi_\pi-\chi_\delta$ (left) and $\chi_{dis}$ (right) obtained at each lattice spacing and $m_\pi=80$ MeV. The factor of $m_s^2/T_c^4$ is used to make the quantities renormalization group invariant and dimensionless. In both plots the open symbols denote direct measurements while the filled symbols slightly shifted horizontally for visibility denote the corresponding results obtained from $\rho$ and $\partial\rho/\partial m_l$. }
	\label{fig:sup_chidisc_V}
\end{figure*}

\begin{figure*}[!thp]
	\includegraphics[width=0.44\textwidth]{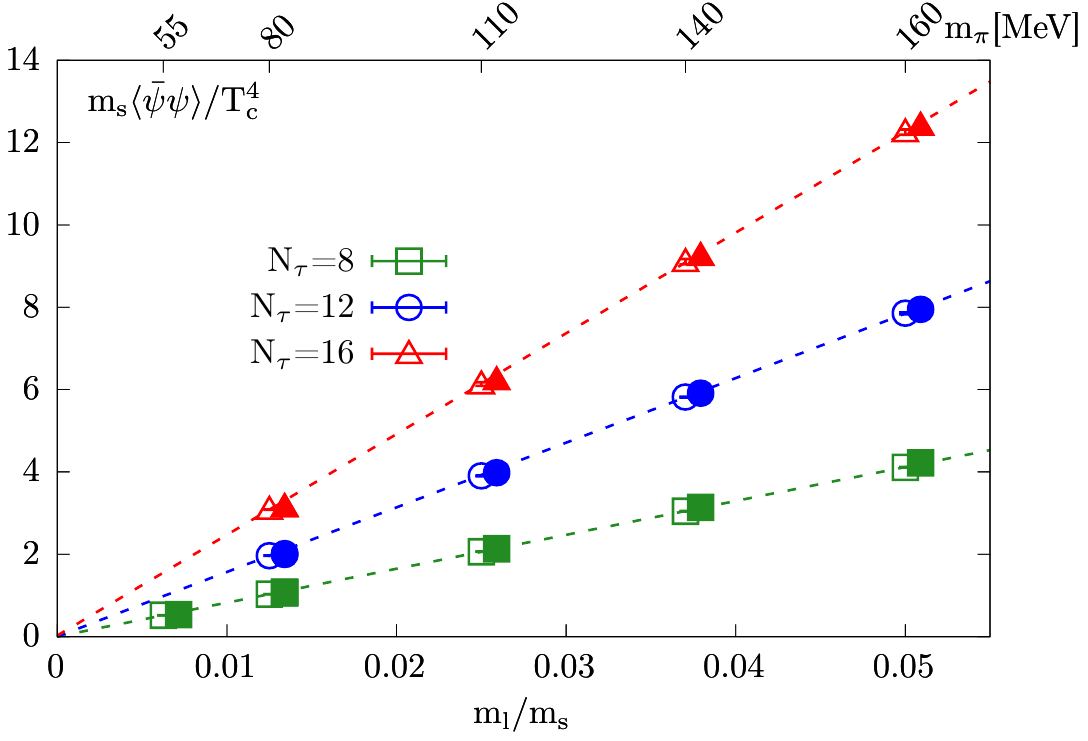}
	\includegraphics[width=0.44\textwidth]{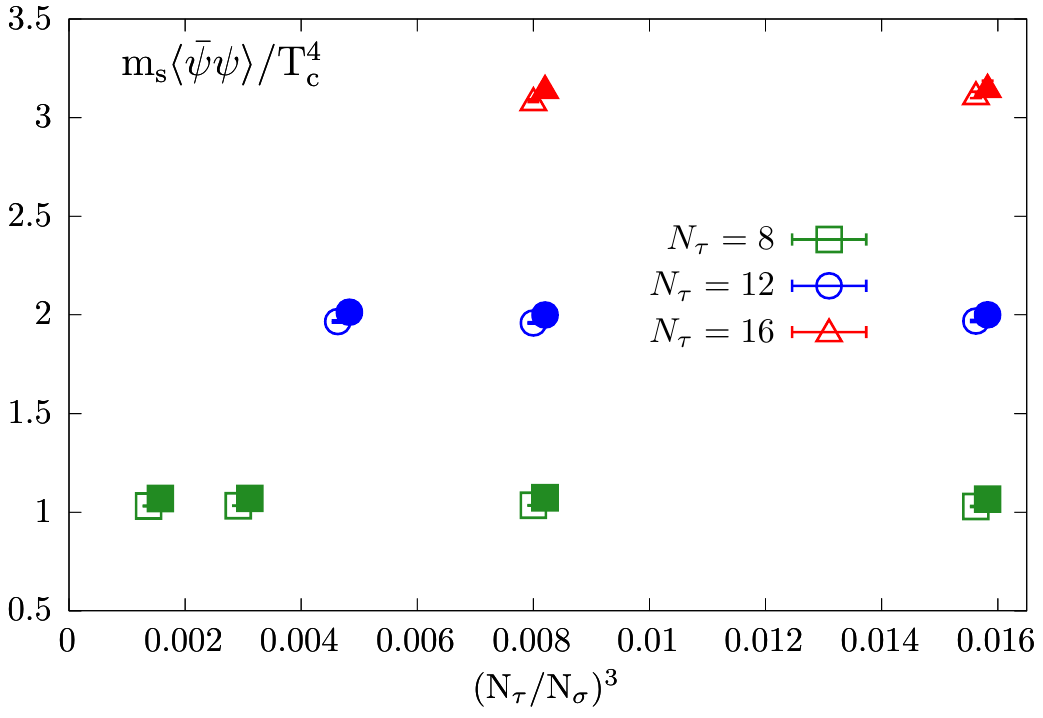}
	\caption{Left: $m_s\pbp/T_c^4$ as a function of light quark mass at different lattice spacings at $m_\pi=80$ MeV obtained from the largest volume available. The dashed lines denote linear fits in quark mass to the directly measured $\pbp$. Right: Volume dependence of $m_s\pbp/T_c^4$ at different lattice spacings with $m_\pi=80$ MeV. In both plots the open symbols denote results obtained from stochastic estimates of the trace of the inverse fermion matrix $M^{-1}$ (\cf~\autoref{eq:sup_pbp_M}) while the corresponding filled symbols slightly shifted horizontally for visibility denote results obtained from $\rho$ via~\autoref{eq:sup_pbp_rho}.}
	\label{fig:sup_su2}
\end{figure*}

\subsubsection{III B2. Reproduction of $\langle\bar{\psi}\psi\rangle$ and $\chi_2$ via $\rho$ and $\pdv[2]{\rho}{m_l}$}

We show the quark mass dependence of light quark chiral condensate in the left panel of~\autoref{fig:sup_su2}. The bin-size of $\rho$, as mentioned in the main text, was fixed by reproducing the value of $\chi_\pi-\chi_\delta$ through our paper. One can see that the chiral condensate computed from the stochastic estimates of $\mathrm{Tr}M^{-1}$ ($\cf$ \autoref{eq:sup_pbp_M}) can be well reproduced by $\rho(\lambda,m_l)$ via ~\autoref{eq:sup_pbp_rho}. The linear fits denoted by the dashed lines give a good description of the data. As seen from the fit results light quark chiral condensates at each lattice spacing vanish in the chiral limit. This is expected in the chiral symmetric phase. Although the data points and fits shown in the left plot are obtained only from the largest $N_\sigma$ available, the volume dependence of chiral condensate is mild as shown in the right panel of~\autoref{fig:sup_su2}.

\begin{figure*}[!thp]
	\includegraphics[width=0.45\textwidth]{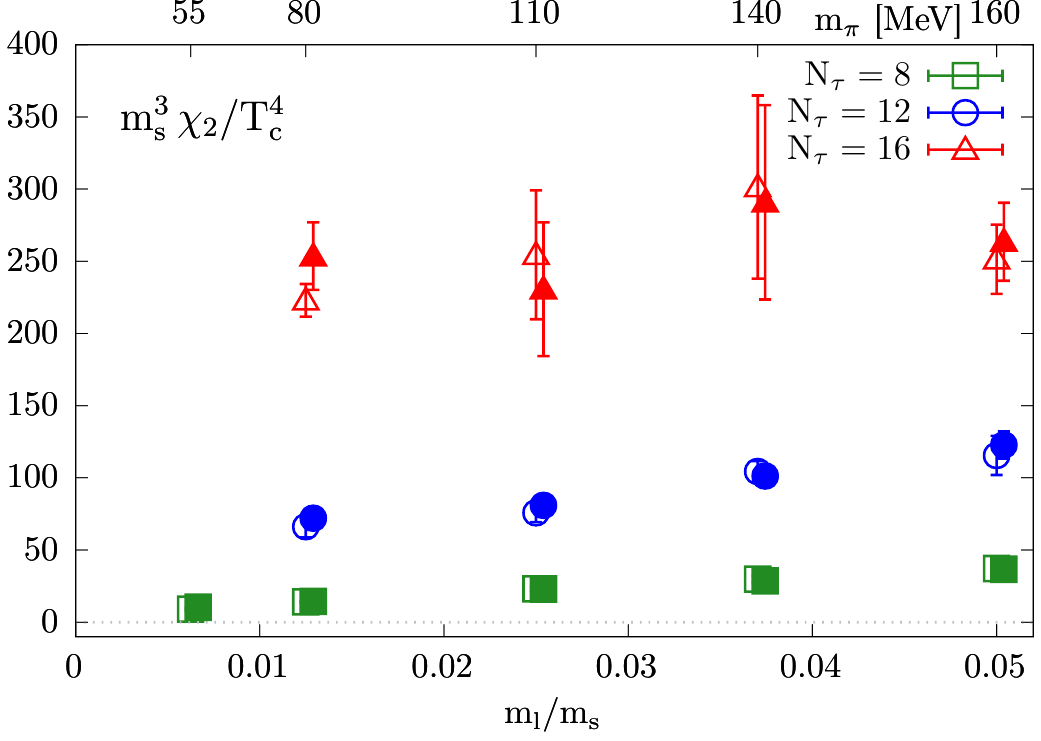}
	\includegraphics[width=0.45\textwidth]{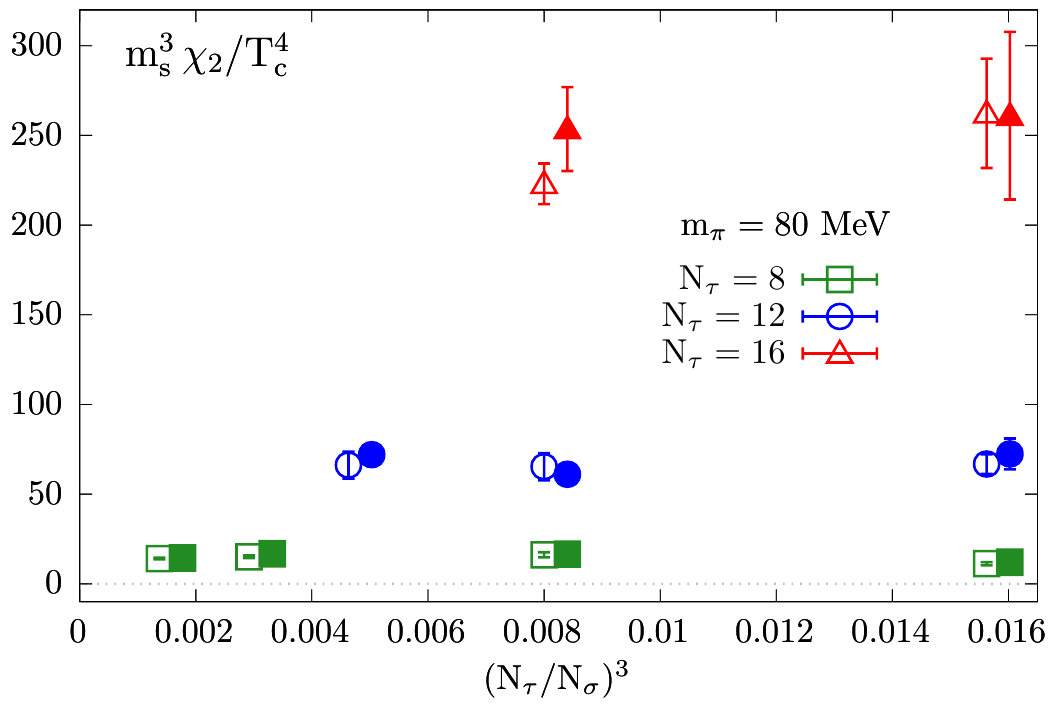}
	\caption{Left: Lattice spacing and quark mass dependences of $m_s^3\chi_2/T_c^4$ obtained from the largest $N_\sigma$ available. Right: Volume dependence of $m_s^3\chi_2/T_c^4$ at $m_\pi=80$ MeV. In both plots the open symbols denote direct measurements from $M^{-1}$ ($\cf$~\autoref{eq:sup_chi2_M}) while the corresponding filled symbols slightly shifted horizontally for visibility denote the corresponding results obtained from $\partial^2\rho/\partial m_l^2$ via \autoref{eq:sup_chi2_rho}. }
	\label{fig:sup_chi2}
\end{figure*}

By using the same bin-size as that used for the reproduction of $\chi_\pi-\chi_\delta$ in the 
numerical integration we observe from \autoref{fig:sup_chi2} that the direct measurement of 
$\chi_2$ ($\cf$ \autoref{eq:sup_chi2_M}) can be well reproduced by 
$\partial^2\rho/\partial m_l^2$ via \autoref{eq:sup_chi2_rho}. It can also be seen that 
this quantity has mild volume dependence at $m_\pi$=80 MeV.

\subsubsection{III B3. Infrared contributions to the two $U(1)_A$ measures}

To check the infrared contribution to the two $\ua$ measures we also introduce an upper cutoff in $\lambda$, i.e. $\lambda_{cut}$ in the integrations of following expressions,
{\begin{align}
\begin{split}
& (\cpi - \cdl) (\lambda_{cut})= \int_0^{\lambda_{cut}} \dd{\lda} \frac { 8 \ml^2\, \rho } { \qty( \lda^2 + \ml^2
	)^2 } \, ,\\
& \cdsc  (\lambda_{cut})= \int_0^{\lambda_{cut} } \dd{\lda} \frac { 4 \ml\,\pdv*{\rho}{\ml} } { \lda^2 +\ml^2  } \,.
\end{split}
\label{eq:sup_susIR}
\end{align}}

\begin{figure*}[!thp]
	\includegraphics[width=0.42\textwidth]{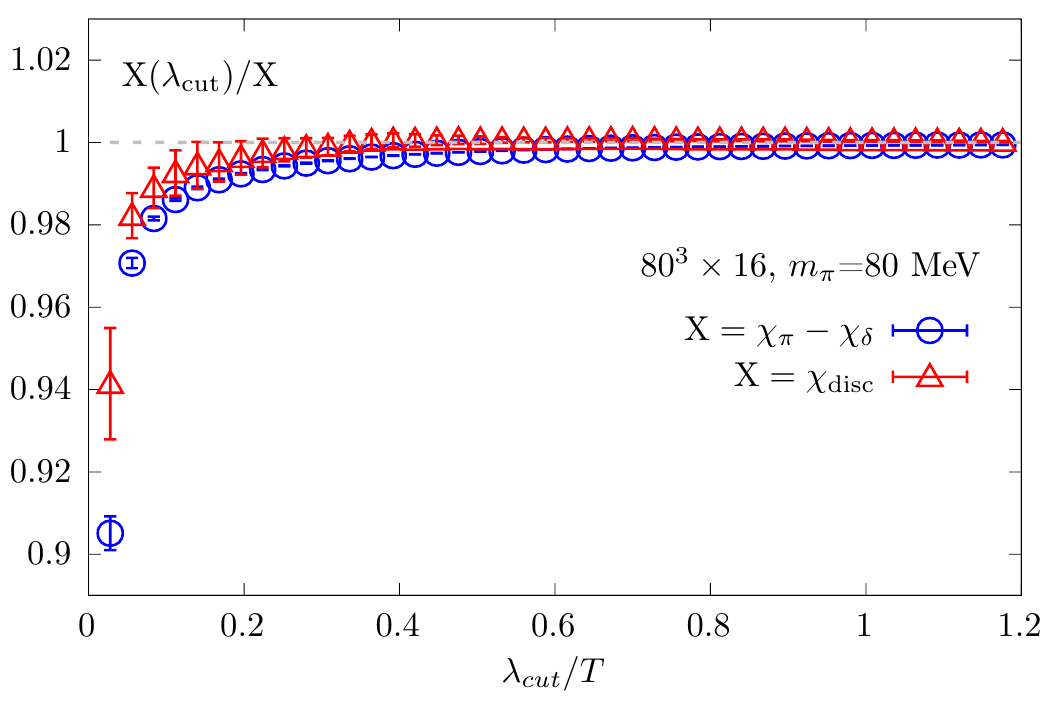}
	\caption{Ratios of $\chi_\pi-\chi_\delta$ and $\chi_{disc}$ obtained from $\rho$ and $\partial\rho/\partial m_l$ with different values of upper limit of the integration $\lambda_{cut}$ to their corresponding values obtained using a complete region of $\lambda$ in the integration.
	}
	\label{fig:sup_chi_IR}
\end{figure*}

We show the $\lambda_{cut}$ dependences of the ratios $(\cpi - \cdl) (\lambda_{cut}) $ and $\chi_{disc} (\lambda_{cut})$ to their corresponding values obtained using the complete $\lambda$ region in the integration in \autoref{fig:sup_chi_IR}. It can be found that the infrared part, i.e. $\lambda/T\lesssim$ 1 of Dirac eigenvalue spectrum gives the dominate contributions to both $\cpi - \cdl$ and $\cdsc$.

\subsection{III C. Supplemental materials to~\autoref{fig:cont_chiral_chi} }
\begin{figure*}[!hpt]
	\includegraphics[width=0.45\textwidth]{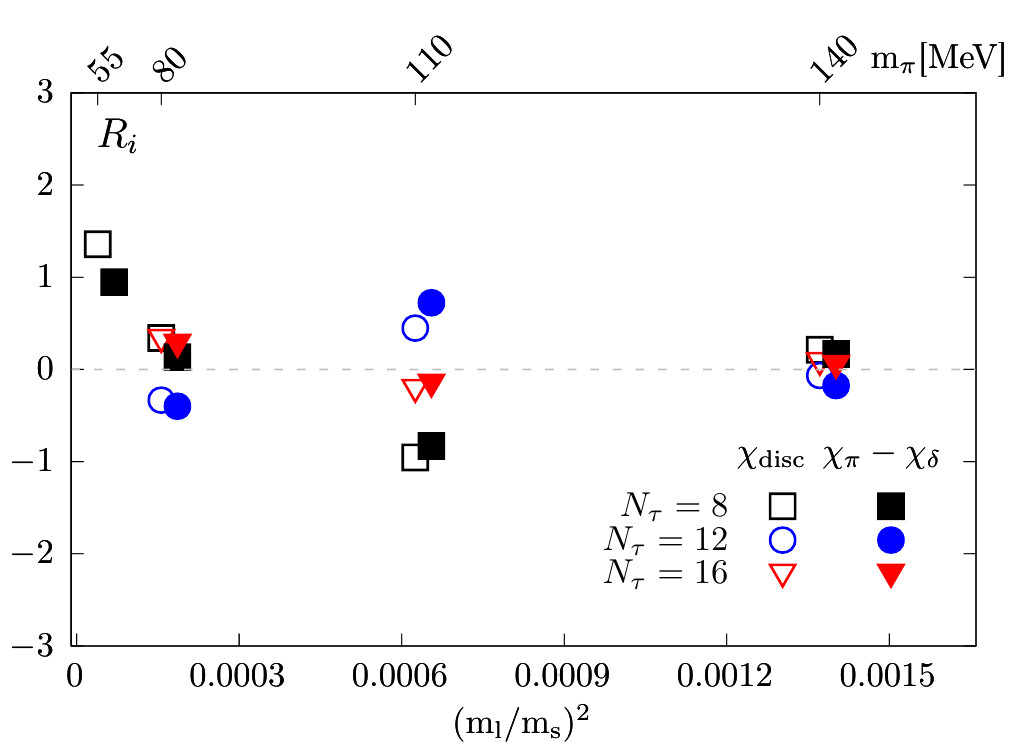}
	\includegraphics[width=0.45\textwidth]{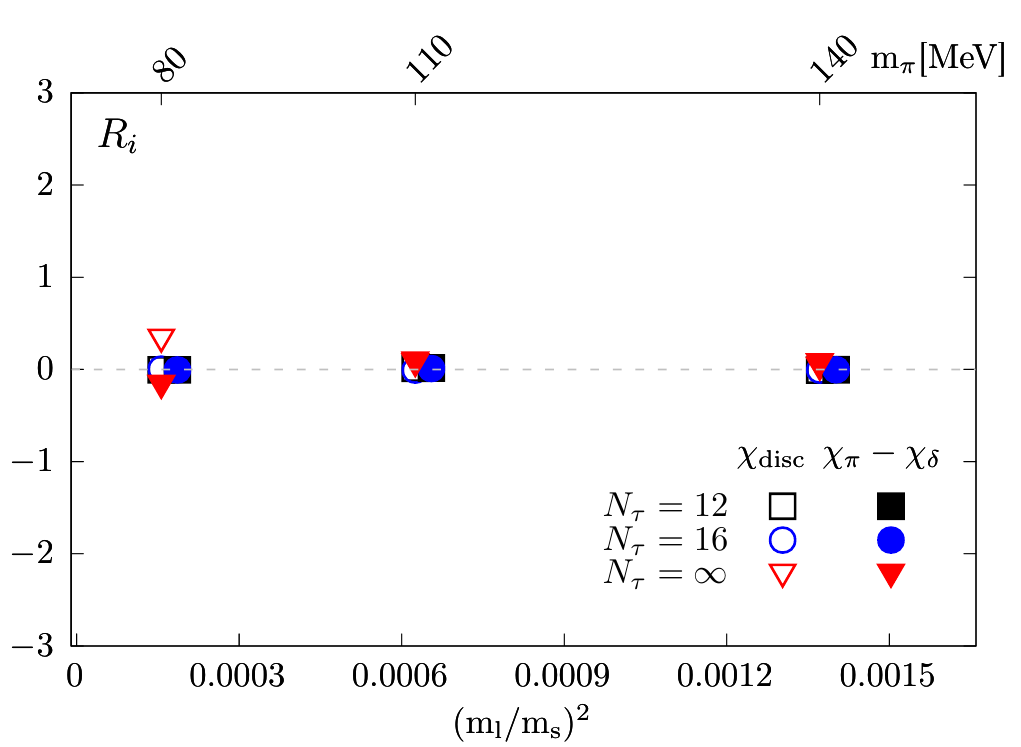}
	\caption{$R_i$ obtained from the joint fit (left) and from the sequential fit (right) with firstly continuum extrapolations and then chiral extrapolations for both $\cdsc$ and $\chi_\pi-\chi_\delta$ performed 
	in~\autoref{fig:cont_chiral_chi} 	in the main material. }
	\label{fig:sup_fits}
\end{figure*}
\begin{figure*}[!hpt]
	\includegraphics[width=0.45\textwidth]{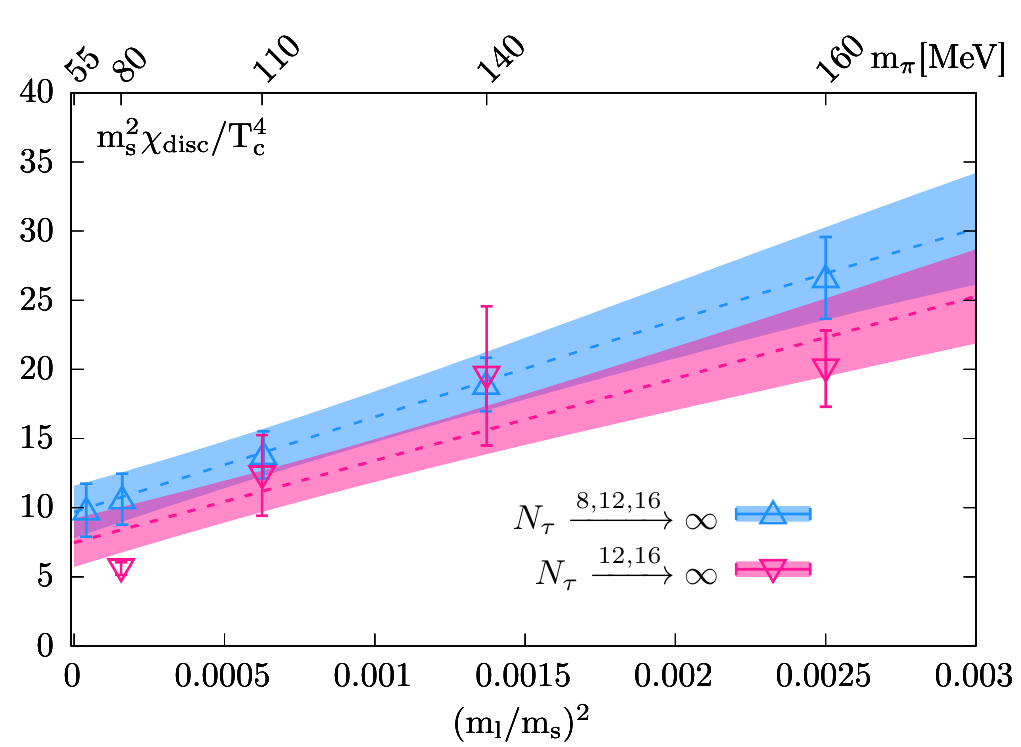}
	\includegraphics[width=0.45\textwidth]{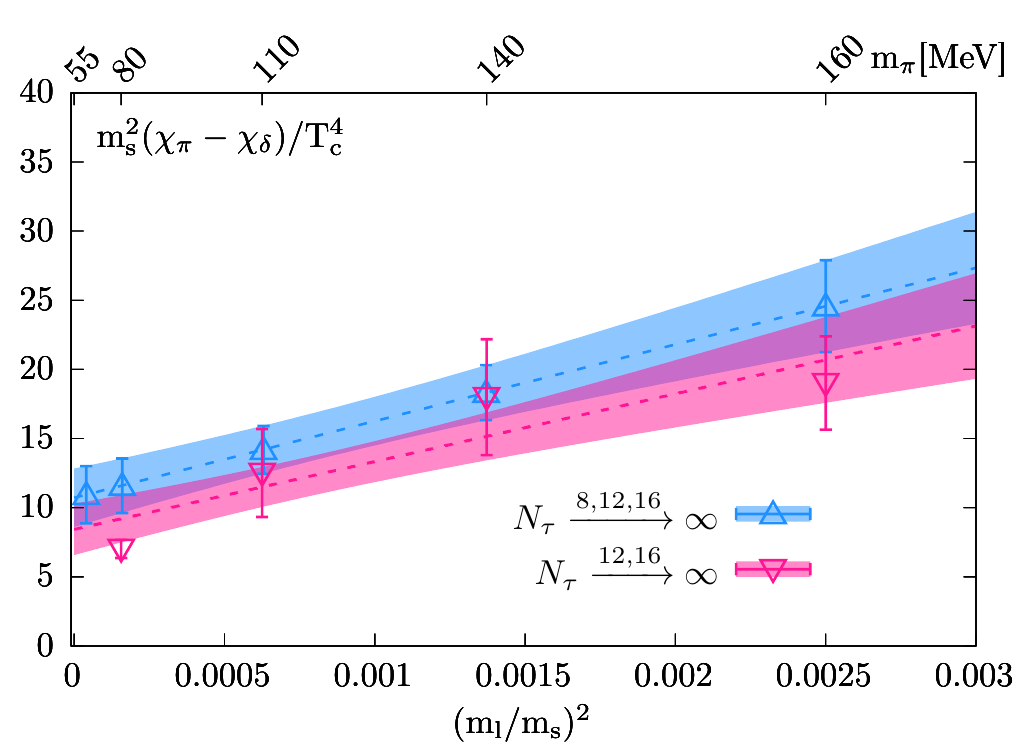}
	\caption{Same as~\autoref{fig:cont_chiral_chi}  in the main material but with data at $m_\pi=$160 MeV included in the extrapolations. }
	\label{fig:fits2}
\end{figure*}
To check the quality of extrapolations we investigate on the following quantity
\begin{align}
R_i = \frac{Y_i - Y_{i,fit}}{\sqrt{\sigma_{Y_i}^2+\sigma^2_{Y_{i,fit}}}}\,.
\label{eq:sup_R}
\end{align}
Here $Y_i$ and $\sigma_{Y_i}$ stand for mean values and Jackknife errors of the data points obtained at each lattice spacing and quark mass, respectively, while $Y_{i,fit}$ and $\sigma_{Y_{i,fit}}$ are corresponding mean values and errors obtained from the fit to data. The integer subscript $i$ runs from 1 to the number of data points used in the fit.

In~\autoref{fig:sup_fits} we show $R_i$ obtained from the extrapolations performed in~\autoref{fig:cont_chiral_chi}  in the main material. I.e. we show $R_i$ obtained from joint fits (left), and from the sequential fit (right) with firstly continuum extrapolations and then chiral extrapolations for both $\cdsc$ and $\chi_\pi-\chi_\delta$. It can be seen that values of $R_i$ scatter around 0.

We perform the same extrapolations as being done in~\autoref{fig:cont_chiral_chi}  in the main material
 but with data at $m_\pi=$160 MeV included. Results for $\chi_{disc}$ and $\chi_\pi-\chi_\delta$ 
 are shown in the left and right panels of
~\textcolor{blue}{Fig.~18}, respectively. It can be seen that the
   two $\ua$ measures remain degenerate within errors in the continuum
	and chiral limit. It can also be observed that both mean values become
	 larger by about 35-73\% while errors remain similar as compared to the 
	 case without $m_\pi=$160 MeV data included in the extrapolations. Thus
	  $\cdsc$ and $\chi_\pi-\chi_\delta$ deviate further away from zero at a (4-5)-$\sigma$ 
	  level. This is expected as that $m_\pi$=160 MeV data at each lattice spacing grows 
	  slower than linearly in quark mass squared ($\cf$~\autoref{fig:Comparison} in the main material) and including it in the extrapolation could bring the values of $\cdsc$ and $\chi_\pi-\chi_\delta$ larger in the chiral limit. To better describe the data at $m_\pi=160$ MeV a fit ans\"atz including higher order corrections in quark mass squared is needed.
	
\end{widetext}
\end{document}